\documentclass[prd,onecolumn,notitlepage,showpacs,preprintnumbers,amsmath,amssymb,nofootinbib,APS,10pt,superscriptaddress]{revtex4-1}

\usepackage{dcolumn}
\usepackage{bm}
\usepackage{xcolor}
\usepackage[utf8]{inputenc}
\usepackage[spanish,english]{babel}
\usepackage{amsmath,amssymb,amsfonts,latexsym,cancel}
\usepackage{graphicx}
\usepackage{color}
\usepackage{soul}
\usepackage{ulem}
\usepackage{hyperref}
\usepackage{amsmath}
\usepackage{slashed}
\usepackage{braket}

\allowdisplaybreaks

\newcommand{\be}{\begin{equation}}
\newcommand{\ee}{\end{equation}}
\newcommand{\bea}{\begin{eqnarray}}
\newcommand{\eea}{\end{eqnarray}}

\begin{document}

\title{{\bf 
Adiabatic regularization for Dirac fields in time-varying electric backgrounds}} % coupled to an electric background in 4 dimensions}} 

\author{Pau Beltr\'an-Palau}\email{pau.beltran@uv.es}
\author{Jose Navarro-Salas}\email{jnavarro@ific.uv.es}
\author{Silvia Pla}\email{silvia.pla@uv.es}

\affiliation{Departamento de Fisica Teorica and IFIC, Centro Mixto Universidad de Valencia-CSIC. Facultad de Fisica, Universidad de Valencia, Burjassot-46100, Valencia, Spain.}

\begin{abstract}

The adiabatic regularization method was originally proposed by Parker and Fulling to renormalize the energy-momentum tensor of scalar fields in  expanding universes.  It can be extended to renormalize the electric current induced by quantized scalar fields in a time-varying electric background. This can be done in a way consistent with gravity if the  vector potential  is considered as a variable of adiabatic order one. Assuming this, we further extend the method to deal with Dirac fields  in four spacetime dimensions. 
This requires a self-consistent ansatz for the adiabatic expansion, in presence of a prescribed time-dependent electric field, which is different from the conventional expansion used for scalar fields. 
Our proposal is consistent, in the massless limit, with the conformal anomaly. We also provide evidence  that our proposed adiabatic expansion for the fermionic modes  parallels the Schwinger-DeWitt adiabatic expansion of the two-point function.   We give the renormalized expression of the electric current and analyze, using numerical and analytical tools, the pair  production induced by a Sauter-type electric pulse. We also analyze the scaling properties of the current for a large field strength.  \\

%In this work we extend the adiabatic regularization method to 4-dimensional 1/2 fields in electric backgrounds. We give the regularized expression of the electric current and analyze the particle production phenomena for the particular case of the Sauter Pulse. We also test the consistency of the method and propose an extended adiabatic regularization method in terms on an arbitrary mass scale $\mu$. \\


 \end{abstract}


\date{\today}
\maketitle
\section{Introduction}\label{Introduction}

The landmark work of Heisenberg and Euler \cite{Heisenberg1936}, motivated by earlier work of Sauter \cite{Sauter}, established the instability of the quantum vacuum under the influence of a prescribed  (slowly varying) electric field. If the field  is sufficiently strong, real electron-positron pairs can be created. This result was reobtained by Schwinger in the modern language of quantum electrodynamics by finding a positive imaginary contribution to the effective action $W$. The quantity $e^{-2 Im \ W}$ represents then the probability that no actual pair creation occurs during the history of the field \cite{Schwinger51}. \\

The  quantum mechanism driving the spontaneous creation of particles by a  gravitational field was discovered  by Parker in the early sixties by studying  quantized fields in an expanding universe. The crucial fact is as follows \cite{parker66}:  creation and annihilation operators evolve, under the influence of the expansion of the universe (or a generic time-varying gravitational field), into a superposition of creation and annihilation operators. During a cosmic period when the expansion factor is almost constant one can  interpret the effect of the gravitational field on the particle number and unambiguously establish  the spontaneous creation of  real particles by the evolving gravitational field. %This idea  played an increasingly important role in gravitational physics. 
Major applications of this remarkable phenomena occurs in the very early universe \cite{Zeldovich1970, Ford1977} and in the vicinity of a collapsing star forming a black hole \cite{hawking74}. 
These pioneer works on particle creation launched the  theory of quantum fields in curved spacetime, as a first step to merge gravity and quantum mechanics within a self-consistent and successful framework \cite{parker-toms, Waldbook, fulling, birrell-davies}.
The underlying machinery  was also  employed  to study  time-varying electromagnetic fields  \cite{narozhnyi1970, Brezin-Itzykson}. In the limit of a slowly varying electric field the Schwinger  result can be recovered. \\

In the gravitational scenario, the most relevant physical observable is the energy-momentum tensor. % It is well known that 
Its vacuum expectation value $\langle T_{\mu\nu} \rangle $ possesses ultraviolet (UV) divergences and has to be regularized and renormalized. In the seventies many methods were proposed  to this end, as explained in the monographs \cite{parker-toms, Waldbook, fulling, birrell-davies}. 
For homogeneous, time-dependent spacetimes  a generic expression for $\langle T_{\mu\nu} \rangle$ was obtained for scalar fields within the so-called adiabatic regularization scheme \cite{parker-fulling, Birrell78, Anderson-Parker, ANOP, delRio-Navarro15}. The adiabatic  method uses  a mode by mode subtraction process, naturally suggested by the definition of a single-particle state in an expanding universe, and in such a  way that  preserves the basic symmetries of the theory. %The adiabatic method is naturally suggested by the definition of single-particle states in an expanding Friedmann-Lemaitre-Robertson-Walker (FLRW) universe, required to  establish the particle creation phenomena in gravitation. 
 Furthermore, the adiabatic  method has been proved to be equivalent to the point-splitting Schwinger-DeWitt renormalization scheme \cite{Birrell78, Anderson-Parker, delRio-Navarro15}. The adiabatic expansion of the field modes parallels  the Schwinger-DeWitt  adiabatic expansion of the Feynman propagator in Friedmann-Lemaitre-Robertson-Walker (FLRW) spacetimes. The advantage of the adiabatic regularization method is that it very efficient to implement numerical computations, and it is widely used in cosmology. 
%This method is widely used in cosmology and it turns out to be very efficient to implement numerical computations. 
  It has been recently extended to spin-$1/2$ fields in FLRW universes \cite{rio1, BFNV}. \\

 %It has also been extended to include an scalar background field \cite{Yukawa}.\\
 
 As  mentioned above, the analysis of particle creation by time-varying electric fields can be carried out using the techniques first proposed to treat curved backgrounds. The electromagnetic field is considered as an external, unquantized background, while the created particles are excitations of the quantized matter field. From the experimental side, this particle production effect is also of special interest since it may not be far from being experimentally detected in high intensity lasers \cite{ELI},  and in beam-beam collisions \cite{Yakimenko19}. %Time varying electric fields interacting with charged electric fields can also create particles. In the limit of constant electric field one recovers the Schwinger effect \cite{Schwinger51}. This effect can may be at the verge of being experimentally detected in lasers of unprecedented intensities \cite{ELI}.  In this scenario, and analogously to the gravitational case, one can consider the electric field as an external classical source, while the created particles can be treated as quantum particles. \\
This effect is also very important in astrophysical \cite{R,Kim:2019joy} and cosmological scenarios \cite{Cosmo1,Scahl,SGF19}, and in nonequilibrium processes induced by strong fields \cite{Dunne}. In this context, the most important physical local expectation value  is the electric current $\langle j_\mu \rangle$, which also possesses ultraviolet divergences and has to be renormalized in a proper way.  Recent discussions on foundational issues related to the particle number density of the created particles, adiabatic invariance, and  unitary evolution can be seen in  \cite{Dunne2,Garay,BFNP}.\\

 Due to the similarities with the gravitational case, it is a good strategy to readapt the adiabatic regularization scheme to the case in which the external background is an electric field. This program was initiated in \cite{Cooper-Mottola89, Kluger91} to study backreaction problems when the matter field is a charged scalar field. It was further extended to treat  charged Dirac fermions \cite{Kluger92}. It was assumed that the adiabatic order of the vector potential $A_\mu$ is 0. The problem was reconsidered for a charged scalar field in \cite{FN} by assuming that the adiabatic order of   $A_\mu$ is $1$, instead of zero.  This new reassignment of the adiabatic order for $A_\mu$ is an unavoidable requirement in presence of a gravitational background. The argument was  reinforced in \cite{FNP, FNP2} on the basis of the covariant conservation of the energy-momentum tensor.  The adiabatic regularization of two-dimensional fermions incorporating the  adiabatic order assignment $1$ for $A_\mu$ has been further reanalyzed in \cite{FN, BFNP, BNP}.
% {\color{blue}This method was also extended to the case of a four dimensional Dirac field by fixing the adiabatic order of the vector potential $A_\mu$ as 0 \cite{Kluger92}, but it has been proved that this renormalization scheme is not appropriate since it fails when including a gravitational background \cite{FNP}. The most proper one consists in considering $A_\mu$ of adiabatic order 1, and this is the scheme that we will use in this work.} 
Other renormalization methods have been  generalized to incorporate an electromagnetic background, as for instance the Hadamard point-splitting method for complex scalar fields \cite{PointS, Had1}, with results in agreement with \cite{FN}. \\

Within the above context, it seems natural to extend   the adiabatic regularization/renormalization  method, with the assumption that $A_\mu$ is of adiabatic order $1$,  to  Dirac fields in  presence  of an electric field background in four spacetime dimensions. This is the main  aim of this work. As stressed above, previous studies in the literature on this problem \cite{Kluger92} assumed that  $A_\mu$ is of adiabatic order $0$.  This extension requires a self-consistent ansatz for the adiabatic expansion of the field modes. We give a proper ansatz,  which cannot be fitted within the conventional Wentzel-Kramers-Brillouin (WKB)-type expansion used for scalar fields \cite{parker-toms, fulling, birrell-davies}. Our extension of the adiabatic method is in agreement with the trace anomaly. Even more, we provide strong evidence that our adiabatic expansion of the field modes parallels the adiabatic Schwinger-DeWitt expansion of the propagator. In addition to the trace anomaly, our adiabatic expansion also reproduces the DeWitt coefficient $E_3$, at sixth adiabatic order.  We carry out the adiabatic renormalization and provide a general expression for the renormalized electric current. We illustrate the power of the method by studying with detail a Sauter-type electric pulse.  \\

The paper is organized as follows. In Sec. \ref{section2} we will describe the status of  adiabatic regularization when a time-varying electric field is part of the background. %We will explain  the  unsatisfactory status of previous analysis in the literature. 
We will give strong reasons for adopting a new viewpoint and reprehend the problem of the adiabatic regularization of  charged $4d$ fermions in time-dependent electric fields. In Sec. \ref{S1} we introduce the basic ingredients of our  ansatz to construct the adiabatic expansion of the four-dimensional fermionic modes coupled to a prescribed time-dependent electric field. %we raise the problem of a quantum 4-dimensional fermionic field interacting with an homogeneous and time-dependent electric background, obtaining the motion equations of the field modes. In Section \ref{AdExp} we propose the adiabatic expansion for the fermionic field modes. 
Sec. \ref{adReg} is devoted to explain the details of the adiabatic renormalization  procedure in this context. In particular, we give a generic and  explicit expression of the renormalized electric current. We also test the consistency of the method and discuss some intrinsic renormalization ambiguities. %contemplate the possibility of performing a modified adiabatic expansion in terms of an arbitrary mass scale $\mu$ and its physical consequences. 
In Sec. \ref{PhysicalApplication} we study the particular case in which the background field is a Sauter-type electric pulse. We analyze the particle production phenomena in terms of the renormalized electric current. We also discuss the scaling properties of the created current. In Sec. \ref{conclusions} we state our main conclusions. Our work is complemented with a series of appendices where we give technical details. We also discuss in Appendix \ref{appendixB} the connection  between the adiabatic  method and the Hadamard renormalization scheme for charged scalar fields. 

 \section{Background and motivation}\label{section2}
%\section{$4d$ charged scalar fields, $2d$ charged fermions, and motivation}\label{section2}

To motivate the main idea of this work it is very convenient to present the status of  the adiabatic regularization method for a charged $4$-dimensional scalar field interacting with a classical,  homogeneous, time-dependent electric background. We will assume that the electric field  is of the form $\vec E=(0,0,E(t))$ with potential vector $A_\mu=(0,0,0,-A(t))$. We will also assume that the  spacetime is described by a FLRW metric of the form $ds^2= dt^2 - a^2(t)d\vec x^2$.  The Klein Gordon equation reads
\bea
(D_\mu D^\mu +m^2+ \xi R)\phi=0\ ,
\eea
where $D_\mu\phi=(\nabla_\mu+iqA_\mu)\phi$ and  $R$ is the Ricci scalar. Since the potential vector $A_\mu$ is homogeneous, one can expand the scalar field in modes as $\phi=\frac{1}{\sqrt{2(2\pi a)^3}}\int d^3k(A_{\vec k} e^{i\vec k\vec x}h_{\vec k}+B_{\vec k}e^{-i\vec k\vec x}h^{*}_{-\vec k})$, where the mode functions $h_{\vec k}(t)$ satisfy
\bea
\ddot h_{\vec k} +(a^{-2} (k_3+qA)^2+ a^{-2} k_\perp^2+m^2 + \sigma)h_{\vec k}=0\ \label{modeS0}
\ , \eea
with $\sigma= (6\xi -3/4)\dot a^2/a^2 +(6\xi -3/2)\ddot a/a$. 
%The scalar modes satisfies the Wronskian condition $wronskian$. 
Once we have obtained the mode equation (\ref{modeS0}), we can make an adiabatic expansion of the field modes. To this end, one can  propose the usual WKB ansatz
\bea
h_{\vec{k}}=\frac{1}{\sqrt{\Omega_{\vec{k}}}}e^{-i\int \Omega_{\vec{k}}(t)dt}\ ,
\eea
where $W_k$ can be expanded adiabatically, in powers of derivatives of $a(t)$ and $A(t)$, as $\Omega_{\vec{k}}=\sum_{n=0}^{\infty}\omega^{(n)}$. \\

The choice of the leading terms $\omega^{(0)}$ is a crucial ingredient to define the adiabatic expansion.  For $A=0$ the proper choice for $\omega^{(0)}$ is $\omega^{(0)}=\omega=\sqrt{\vec{k}^2/a^2+m^2}$. This defines the conventional adiabatic expansion for a scalar field, as first introduced in the pioneer works \cite{parker-fulling}. When the background spacetime is Minkowski $a=1$, the choice proposed in \cite{Cooper-Mottola89} was  $\omega^{(0)}=\omega=\sqrt{(\vec{k} - q\vec A)^2+m^2}$. This choice assumes that $A(t)$ should be treated as a variable of zero adiabatic order, like $a(t)$. As noted in \cite{FN, FNP}, this choice runs into difficulties in  presence of a gravitational background. It was proposed  in \cite{FN, FNP} that the leading term should be maintained as  $\omega^{(0)}=\omega=\sqrt{\vec{k}^2/a^2+m^2}$, even in the presence of an electromagnetic field.   This means that $A(t)$ must be considered as a variable of adiabatic order $1$, like $\dot a$. Hence $\dot A$ is  of adiabatic order $2$, etc.   \\

 Next to leading order terms can be obtained recursively from (\ref{modeS0}). The adiabatic expansion allows us to regularize the observables performing adiabatic subtractions. %, that is, subtracting the first terms of the adiabatic expansion to our physical observables obtaining finite and meaningful results. The first terms of the adiabatic expansion capture all potential divergences. 
Since the first terms of the adiabatic expansion capture all potential ultraviolet divergences, one can subtract them, %to physical observables
  obtaining finite and meaningful results. 
With this method, we obtain the following vacuum expectation value of the two point function

%Since the first terms of the adiabatic expansion capture all potential divergences, we can subtract them to our physical observables,  obtaining finite and meaningful results.
%(revisar factor 1/2 en $T_{00}$)
%\subsubsection{2Dimensiones}
%\bea
%&&\langle \phi \phi^\dagger\rangle_{ren}=q \int_{-\infty}^{\infty} \frac{dk}{2 \pi } \left [|h_k|^2-\frac{1}{\omega} \right ] \label{renjgrav} \ .
%\eea
%
%\bea
%&&\langle j^{x}\rangle_{ren}=q \int_{-\infty}^{\infty} \frac{dk}{2 \pi } \left [\left(k-q A\right) |h_k|^2-\frac{k}{\omega}+\frac{m^2 q A}{\omega^3} \right ] \label{renjgrav} \ .
%\eea
%\be
%\langle T_{00} \rangle_{ren}= \frac{1}{4\pi }\int_{-\infty}^{\infty} dk \left [|\dot{h}_k|^2+ m^2|h_k|^2+(k-q A)^2 |h_k|^2 -2\omega + \frac{2 k q A}{\omega} -\frac{m^2 q^2 A^2}{ \omega^3}\right ].\label{scden}
%\ee
%\subsubsection{4Dimensiones}
%\bea
%\langle \phi \phi^\dagger  \rangle_{ren}&=&\frac{ 1}{2(2 \pi)^{3}} \int d^3k\left[|h_{\vec{k}}|^2-\langle \phi \phi^\dagger  \rangle_k^{(0-2)}\right] \label{phiphiA}\\
%\langle j^{3}\rangle_{ren}&=&\frac{ q}{(2 \pi)^{3}} \int d^3k
%\Big[k_3 |h_{\vec{k}}|^2-\Big(k_3 |h_{\vec{k}}|^2\Big)^{(0-3)}\Big]-qA\langle \phi \phi^\dagger  \rangle_{ren}\label{jscalars}\\
%\langle T_{00} \rangle_{ren}&=& \frac{ 1}{2(2 \pi)^{3}} \int d^3k\Big[ |\dot{h}_k|^2+ \kappa^2|h_k|^2+(k-q A)^2 |h_k|^2-\langle T_{00} \rangle_k^{(0-4)} \Big]\\
%\partial_t^2\langle \phi \phi^\dagger  \rangle_{ren}&=&\frac{ 1}{(2 \pi)^{3}} \int d^3k\Big[|\dot h_{\vec{k}}|^2-(\kappa^2+(k-qA)^2)| h_{\vec{k}}|^2-\partial_t^2 \langle \phi \phi^\dagger  \rangle_k^{(0-2)}\Big]
%\eea
\bea
\langle \phi ^\dagger \phi  \rangle_{ren}&=&\frac{ 1}{2(2 \pi a(t))^{3}} \int d^3k\left[|h_{\vec{k}}|^2-\langle \phi ^\dagger \phi \rangle_{\vec{k}}^{(0-2)}\right]\ , \label{phiphiA}
\eea
where $\langle \phi^\dagger  \phi \rangle_{\vec{k}}^{(0-2)}=\sum_{n=0}^2(\Omega_{\vec{k}}^{-1})^{(n)}$. \\

As stressed in the introduction, the adiabatic expansion of the field modes translates into an adiabatic expansion of the two-point function. The latter turns out to be equivalent to the Schwinger-DeWitt expansion of the Feynman propagator $\langle T \phi^{\dagger} (x) \phi(x')\rangle$.  We will show this explicitly at the coincident limit $x' \to x$ at fourth and sixth adiabatic orders.  At fourth adiabatic order the corresponding momentum integral is finite and we get
\bea \langle \phi ^\dagger \phi  \rangle^{(4)} = \frac{ 1}{2(2 \pi a)^{3}} \int d^3k\langle \phi ^\dagger \phi \rangle_{\vec{k}}^{(4)}=  \frac{1}{16\pi^2 m^2}\biggl(\frac{36 \dot{a}^2 \xi ^2 \overset{\text{..}}{a}}{a^3}-\frac{17 \dot{a}^2 \xi 
   \overset{\text{..}}{a}}{a^3}+\frac{29 \dot{a}^2 \overset{\text{..}}{a}}{15
   a^3}+\frac{18 \xi ^2 \overset{\text{..}}{a}^2}{a^2}-\frac{5 \xi 
   \overset{\text{..}}{a}^2}{a^2}+\frac{3 \overset{\text{..}}{a}^2}{10 a^2}+\\ \nonumber 
   +\frac{18
   \dot{a}^4 \xi ^2}{a^4}-\frac{6 \dot{a}^4 \xi }{a^4}+\frac{a^{(4)} \xi
   }{a}+\frac{\dot{a}^4}{2 a^4}-\frac{a^{(4)}}{5 a}+\frac{\dot{A}^2 q^2}{6 a^2}+\frac{3
   a^{(3)} \dot{a} \xi }{a^2}-\frac{3 a^{(3)} \dot{a}}{5 a^2}\biggr) \ . \eea
It is not difficult to check that the above result can be reexpressed in the following covariant form
\be \langle \phi ^\dagger \phi  \rangle^{(4)} =  \frac{E_2}{16\pi^2 m^2}  \ , \ee
where $E_2$ matches exactly the  DeWitt coefficient \cite{DeWittbook}
\bea \label{E20} E_2 = -\frac{1}{30} \Box R + \frac{1}{72} R^2 -\frac{1}{180} R^{\mu\nu}R_{\mu\nu} + \frac{1}{180} R^{\mu\nu\rho\sigma}R_{\mu\nu\rho\sigma} - \frac{q^2}{12} F^{\mu\nu}F_{\mu\nu} 
+ \frac{1}{2} Q^2 - \frac{1}{6} RQ + \frac{1}{6}\Box Q \ . \eea
In the above expression $Q$ is given by $Q=\xi R$. We note that, to obtain  equivalence with the Schwinger-DeWitt proper-time method it has been essential to assume that $A_\mu$ is of adiabatic order $1$, $F_{\mu\nu}$ of adiabatic order $2$, etc. With the  zero  adiabatic order assignment for $A_\mu$ one obtains a noncovariant and ill-defined expression for $\langle \phi ^\dagger \phi  \rangle^{(4)}$. With our proposed leading order choice for  $\omega^{(0)}=\omega=\sqrt{\vec{k}^2/a^2+m^2}$ we find equivalence with the Schwinger-DeWitt expansion at very nontrivial higher orders. For instance, our calculation for $\langle \phi ^\dagger \phi  \rangle^{(6)}$ gives

 \bea &&\langle \phi ^\dagger \phi  \rangle^{(6)} = \frac{ 1}{2(2 \pi a)^{3}} \int d^3k\langle \phi ^\dagger \phi \rangle_{\vec{k}}^{(6)} \nonumber \\
 &=&-\frac{108 \dot{a}^4 \xi ^3
   \overset{\text{..}}{a}}{a^5}+\frac{96 \dot{a}^4
   \xi ^2 \overset{\text{..}}{a}}{a^5}-\frac{217
   \dot{a}^4 \xi  \overset{\text{..}}{a}}{10
   a^5}+\frac{197 \dot{a}^4
   \overset{\text{..}}{a}}{140 a^5}-\frac{108
   \dot{a}^2 \xi ^3
   \overset{\text{..}}{a}^2}{a^4}+\frac{75
   \dot{a}^2 \xi ^2
   \overset{\text{..}}{a}^2}{a^4}-\frac{221
   \dot{a}^2 \xi  \overset{\text{..}}{a}^2}{10
   a^4}+\frac{159 \dot{a}^2
   \overset{\text{..}}{a}^2}{70
   a^4}-\frac{\dot{A}^2 \xi  q^2
   \overset{\text{..}}{a}}{a^3}+ \nonumber \\
   &&+\frac{\dot{a}
   \dot{A} q^2 \overset{\text{..}}{A}}{90
   a^3}+\frac{23 \dot{A}^2 q^2
   \overset{\text{..}}{a}}{90 a^3}-\frac{36 \xi ^3
   \overset{\text{..}}{a}^3}{a^3}-\frac{24 a^{(3)}
   \dot{a} \xi ^2
   \overset{\text{..}}{a}}{a^3}+\frac{12 \xi ^2
   \overset{\text{..}}{a}^3}{a^3}+\frac{133 a^{(3)}
   \dot{a} \xi  \overset{\text{..}}{a}}{10
   a^3}+\frac{3 \xi  \overset{\text{..}}{a}^3}{5
   a^3}-\frac{103 a^{(3)} \dot{a}
   \overset{\text{..}}{a}}{60 a^3}-\frac{14
   \overset{\text{..}}{a}^3}{45 a^3}-\frac{q^2
   \overset{\text{..}}{A}^2}{20 a^2}- \nonumber \\
   &&-\frac{6
   a^{(4)} \xi ^2
   \overset{\text{..}}{a}}{a^2}+\frac{17 a^{(4)}
   \xi  \overset{\text{..}}{a}}{10 a^2}-\frac{43
   a^{(4)} \overset{\text{..}}{a}}{420
   a^2}-\frac{36 \dot{a}^6 \xi ^3}{a^6}+\frac{6
   \dot{a}^6 \xi ^2}{a^6}+\frac{\dot{a}^6 \xi
   }{a^6}-\frac{a^{(6)} \xi }{10
   a}-\frac{\dot{a}^6}{6 a^6}+\frac{3 a^{(6)}}{140
   a}-\frac{\dot{a}^2 \dot{A}^2 \xi 
   q^2}{a^4}+\frac{37 \dot{a}^2 \dot{A}^2 q^2}{180
   a^4}- \nonumber \\
   &&-\frac{\dot{A} A^{(3)} q^2}{15 a^2}-\frac{2
   a^{(5)} \dot{a} \xi }{5 a^2}+\frac{3 a^{(5)}
   \dot{a}}{35 a^2}-\frac{6 a^{(3)} \dot{a}^3 \xi
   ^2}{a^4}-\frac{6 a^{(4)} \dot{a}^2 \xi
   ^2}{a^3}+\frac{7 a^{(3)} \dot{a}^3 \xi }{10
   a^4}+\frac{31 a^{(4)} \dot{a}^2 \xi }{10
   a^3}+\frac{13 a^{(3)} \dot{a}^3}{140
   a^4}-\frac{23 a^{(4)} \dot{a}^2}{60 a^3}- \nonumber \\
   &&-\frac{3
   \left(a^{(3)}\right)^2 \xi ^2}{a^2}+\frac{7
   \left(a^{(3)}\right)^2 \xi }{10
   a^2}-\frac{\left(a^{(3)}\right)^2}{42 a^2}   \ , \eea
   where $a^{(n)}$ refers to $d^n a/d t^n$. 
The result turns out to be proportional to the corresponding DeWitt coefficient of sixth adiabatic order $E_3$ \cite{Gilkey, Vassilevich}. 
The covariant expression is given in the Appendix \ref{appendixA0}.\\

We stress again that it has been  crucial for obtaining the above results the choice $\omega^{(0)}=\omega=\sqrt{\vec{k}^2/a^2+m^2}$, instead of $\omega^{(0)}=\omega=\sqrt{(\vec{k}-q\vec A)^2/a^2+m^2}$. For completeness, a comparison of the above formulation of the adiabatic regularization with the Hadamard renormalization scheme is given in the Appendix \ref{appendixB}.  \\
%{\color{blue} Finally, we want to remark that the  consistency for the adiabatic order assignment $1$ for  $A_\mu$ does not require the presence of gravity. Even in Minkowski space, the presence of an additional interaction like a Yukawa coupling  of the form $\Phi^2 \phi \phi^\dagger$, demands  that the adiabatic order of the external scalar field $\Phi$, as well as $A_\mu$,  are fixed to be  $1$. This can read directly from the general expression for $E_2$. In Minkowski space $Q = \Phi^2$ must be of adiabatic order $2$ and $\Box Q$ of adiabatic order $4$. Hence $F_{\mu\nu}$ should also be of adiabatic order $2$.  In  presence of the Yukawa interaction  the adiabatic expansion of the field modes is also equivalent to the Schwinger-DeWitt adiabatic expansion of the Feynman propagator \cite{FNP20}. } \\

\subsection{Adiabatic regularization for fermions in two-dimensions}

To reinforce the previous analysis, and prior to face the adiabatic regularization of charged fermions in four spacetime dimensions, it is also convenient to consider the problem for a charged Dirac  field in two-dimensions. We will follow \cite{FN, FNP} and compare the results with the pioneer analysis in \cite{Kluger92}. The comparison will allow us to understand why it has been necessary to reprehend the problem, as already stressed above.  \\

The quantum field satisfies the Dirac equation $(i \underline \gamma^{\mu}D_{\mu}-m)\psi=0$, where $D_{\mu} \equiv \nabla_{\mu} -\Gamma_\mu-i q A_{\mu}$ and $\Gamma_\mu$ is the spin connection. The curved space Dirac matrices satisfy the anticommutation relations $\{\underline \gamma^{\mu},\underline \gamma^{\mu}\}=2g^{\mu\nu}$. We assume a  homogeneous, time-dependent electric background $E(t)$, with associated potential vector $A_\mu=(0,-A(t))$. The metric is also assumed of the FLRW form $ds^2= dt^2 - a^2(t)dx^2$. One can expand the Dirac field as $\psi=\int_{-\infty}^\infty dk[B_ku_k(t,x)+D^{\dagger}v_k(x,t)]$, where the two independent spinor solutions can be written as
\bea
 u_{k}(t, x)=\frac{e^{ikx}}{\sqrt{2\pi a}} \scriptsize \left( {\begin{array}{c}
 h^{I}_k(t)   \\
 -h^{II}_k (t) \\
 \end{array} }\right)\ , \hspace{1cm}
  v_{k}(t, x)=\frac{e^{-ikx}}{\sqrt{2\pi a}} \scriptsize \left( {\begin{array}{c}
 h^{II*}_{-k} (t)  \\
 h^{I*}_{-k}(t)  \\
 \end{array} } \right)
 \label{spinorde}
\ .\eea

The classical electric field satisfies the semiclassical Maxwell equations $\nabla_\mu F^{\mu \nu}=-q\langle\bar \psi \underline \gamma^\nu \psi \rangle_{ren} =\langle j^\nu \rangle_{ren}$, which in our system turns out to be a single equation $\dot E=-\langle j^x \rangle_{ren}$. In this scenario the adiabatic rules are univocally fixed: $a(t)$ has to be considered of adiabatic order 0,  the  energy-momentum tensor must be regularized up to the second adiabatic order and the electric current must to be regularized up to the first adiabatic order. The adiabatic subtractions required to regularize the electric current $\langle j^x \rangle$
will be different depending on the adiabatic order that we choose for the background field $A(t)$, i.e %: we can consider $A(t)$ of adiabatic order 0 or of adiabatic order 1. % Depending on the choice of the adiabatic order of $A(t)$ we have two possibilities
%In the first case one obtains the following expression for the renormalized electric current
\bea
\langle j^x\rangle_{ren}^{A\sim O(0)}&=&q\int \frac{dk}{2\pi a} \left(|h_k^{II}|^2-|h_k^{I}|^2-\frac{k+qA}{a\sqrt{(k+qA)^2/a^2+m^2}}\right) \label{curr0} \ , \\
\langle j^x\rangle_{ren}^{A\sim O(1)}&=&q\int \frac{dk}{2\pi a} \left(|h_k^{II}|^2-|h_k^{I}|^2-\frac{k}{a\omega}-\frac{m^2qA}{a\omega^3}\right) \ , \label{curr1}
\eea
where $\omega=\sqrt{k^2/a^2+m^2}$. In \eqref{curr0} we have considered $A$ of adiabatic order zero, while in \eqref{curr1} we have considered it of adiabatic order one. % Note 
One can check that the subtractions obtained in the first case are the same to the ones obtained in \cite{Kluger92} for $a=1$.  Although it can be proven \cite{FNP2} that these two choices are equivalent when $a=1$,  in the sense that  $\bigtriangleup \langle j^x\rangle_{ren}=\langle j^x\rangle_{ren}^{A\sim O(0)}-\langle j^x\rangle_{ren}^{A\sim O(1)}=0$, they are in general nonequivalent. %Proceeding similarly, one can obtain the expressions for the energy-momentum tensor in both schemes.
We can see how gravity breaks this equivalence.  In the second case we can easily see that the energy density  is covariantly conserved 
%\be \nabla_{\mu}\langle T^{\mu0}\rangle_{ren}+ \nabla_\mu T^{\mu0}_{elec}=\frac{ \dot{A}}{a}\left(\frac{ \ddot{A}}{a}-\frac{ \dot{A}\dot{a}}{a^2}- \langle j^x \rangle_{ren} \right)=0 \ . \ee
\be \nabla_{\mu}\langle T^{\mu0}\rangle_{ren}+ \nabla_\mu T^{\mu0}_{elec}=E\left(\dot E + \langle j^x \rangle_{ren} \right)=0 \ . \ee
But, when we consider $A(t)$ of adiabatic order 0, the conservation does not hold any more, and one finds $\nabla_{\mu}\langle T^{\mu0}\rangle_{ren}+ \nabla_\mu T^{\mu0}_{elec}\sim E(t) \langle j^x \rangle^{(2)} $, where $\langle j^x \rangle^{(2)}$ is the subtraction term of adiabatic order two, which cannot be properly absorbed into the definition of the electric current.\\

Moreover, only when $A$ is considered of adiabatic order $1$ the adiabatic expansion of the field modes turns out to be equivalent to the Schwinger-DeWitt expansion of the two-point function. For instance, the adiabatic expansion of the two-point function at coincidence is found to be (at second and fourth adiabatic order)

\be \langle \bar \psi \psi \rangle^{(2)} = \frac{1}{4\pi m}\left(\frac{\ddot a}{3 a}\right)=-\frac{ \operatorname{tr}  E_1}{4\pi m} \ , \ee
\be \langle \bar \psi \psi \rangle^{(4)} = \frac{1}{4\pi m^3}\left(-\frac{\dot{a}^2 \ddot{a}}{30 a^3}+\frac{\ddot{a}^2}{15 a^2}-\frac{a^{(4)}}{30 a}+\frac{2 \dot{A}^2 q^2}{3 a^2}+\frac{a^{(3)} \dot{a}}{30 a^2}\right)=-\frac{\operatorname{tr} E_2}{4\pi m^3}  \ , \ee
where $E_1$ and $E_2$ are the corresponding DeWitt coefficients. They  are given, in the covariant form,  by \cite{parker-toms, Gilkey, Vassilevich}
\be E_1=\frac{1}{6}R I -Q \ , \ee
\be E_2 =( -\frac{1}{30} \Box R + \frac{1}{72} R^2 -\frac{1}{180} R^{\mu\nu}R_{\mu\nu} + \frac{1}{180} R^{\mu\nu\rho\sigma}R_{\mu\nu\rho\sigma})I + \frac{1}{12} W^{\mu\nu}W_{\mu\nu} 
+ \frac{1}{2} Q^2 - \frac{1}{6} RQ + \frac{1}{6}\Box Q \ , \ee
where $Q=\frac{1}{4}R I-\frac{i}{2}qF_{\mu\nu}\underline \gamma^\mu\underline \gamma^\nu$ and $W_{\mu\nu}=-i q F_{\mu\nu} I-\frac14 R_{\mu\nu\rho\sigma} \underline \gamma^\rho\underline\gamma^\sigma$.\\

The above arguments make it necessary to reconsider the problem of adiabatic regularization for fermions in time-varying electric backgrounds in four dimensions. We will adopt the view of considering $A_\mu$ of adiabatic order $1$, as advocated in \cite{FN, FNP, BNP, BFNP},  and in contrast to the view adopted in \cite{Kluger92}.  The main  reasons, as exposed above, are (i) expected agreement with the Schwinger-DeWitt adiabatic  expansion of the two-point function at coincidence; (ii) consistency with the covariant conservation of the energy-momentum tensor when gravity is turned on. We think these are convincing  arguments to go further with our proposed approach. For simplicity we will restrict our analysis to Minkowski spacetime.

\section{$4$d Dirac fields: mode equations, ansatz and adiabatic expansion} \label{S1}
Let us consider a massive 4-dimensional  spinor field $\psi$ interacting with a prescribed electric field. The corresponding Dirac equation  reads
\bea
(i \gamma^{\mu}D_{\mu}-m)\psi=0\label{diraceqm} \  ,
\eea
where $D_{\mu} \equiv \partial_{\mu} -i q A_{\mu}$ and $\gamma^{\mu}$ are the (flat-space) Dirac matrices satisfying the anticommutation relations $\{\gamma^{\mu},\gamma^{\nu}\}=2\eta^{\mu\nu}$.
%For convenience, we will use the Weyl representation (with $\gamma^5 \equiv \gamma^0\gamma^1$)
%\bea
%\gamma^0 = \scriptsize
%\left( {\begin{array}{cc}
% 0 & I  \\
% I& 0  \\
% \end{array} } \right),\hspace{0.5cm} 
%\gamma^i = \scriptsize \left( {\begin{array}{cc}
% 0 & \sigma^i  \\
% -\sigma^i& 0  \\
% \end{array} } \right), \hspace{0.5cm} \gamma^5 = \scriptsize \left( {\begin{array}{cc}
% -I & 0  \\
% 0& I \\
% \end{array} } \right) \nonumber
% \ . \eea
%\\
We consider $\psi$ as a quantized Dirac field, while the electromagnetic field is assumed to be a classical and spatially homogeneous field $\vec{E}(t)=(0,0,E(t))$. It is very convenient to choose a gauge such that only the $z$-component of the vector potential is nonvanishing: $A_{\mu}= (0, 0,0,-A(t))$, where $E(t) = -\dot A(t)$. \\

To prepare things to propose a consistent ansatz for the adiabatic expansion of the field modes %For our purposes 
it is very  important to transform the Dirac field as $\psi'=U\psi$, where $U$ is the unitary operator $U=\frac{1}{\sqrt{2}}\gamma^0(I-\gamma^3)$, which verifies $U=U^\dagger=U^{-1}$. This transformation will allow us to express the Dirac field in terms of only two time-dependent functions [see \eqref{ansatz}]. The field $\psi'$ obeys the Dirac equation for the transformed matrices $\gamma'^{\mu}=U\gamma^{\mu}U^{\dagger}$, namely: $\gamma'^{0}=\gamma^3\gamma^0$, $\gamma'^{1}=-\gamma^3\gamma^1$, $\gamma'^{2}=-\gamma^3\gamma^2$, $\gamma'^{3}=-\gamma^3$. Substituting them in the Dirac equation we easily get
\be
\left[\gamma^0\partial_0-\gamma^1\partial_1-\gamma^2\partial_2-\partial_3-iqA(t)-im\gamma^3\right]\psi'=0 \ .
\ee
Expanding the field in %{\color{red}positive-frequency} 
Fourier modes, $\psi'(t, \vec x)=\int\frac{d^3\vec{k}}{(2\pi)^{3/2}}\psi'_{\vec k}(t)e^{i\vec k\vec x}$, we obtain the following equation
\be \label{dirac2}
\left[\partial_0-i\gamma^0(k_1\gamma^1+k_2\gamma^2+m\gamma^3)-i(k_3+qA(t))\gamma^0\right]\psi'_{\vec k}(t)=0 \ ,
\ee
where $\vec k \equiv (k_1, k_2, k_3)$. 
The form of the above equation allows us to reexpress the field in terms of two-component spinors as follows
\be \label{ansatz}
\psi'_{\vec k,\lambda}(t)=\left( {\begin{array}{c}
h_{\vec k}^{I}(t)\eta_{\lambda}(\vec k) \\
h_{\vec k}^{II}(t)\lambda\eta_{\lambda}(\vec k)\\	
\end{array} }\right) \ ,
\ee
where
%$\vec\kappa=(k_1,k_2,m)$, $\kappa=\sqrt{k_1^2+k_2^2+m^2}$, and 
$\eta_\lambda$ with $\lambda=\pm1$ form an orthonormal basis of two-spinors ($\eta_\lambda^\dagger\eta_{\lambda'}=\delta_{\lambda,\lambda'}$) verifying $\frac{k_1\sigma^1+k_2\sigma^2+m\sigma^3}{\sqrt{k_1^2+k_2^2+m^2}}\eta_{\lambda}=\lambda\eta_\lambda$. Their explicit expressions are
\bea
\eta_{+1}(\vec k)=\frac{1}{\sqrt{2\kappa(\kappa+m)}}\left( {\begin{array}{c}
\kappa+m \\
k_1+ik_2\\	
\end{array} }\right) \ , \nonumber \\
\eta_{-1}(\vec k)=\frac{1}{\sqrt{2\kappa(\kappa+m)}}\left( {\begin{array}{c}
-k_1+ik_2\\	
\kappa+m \\
\end{array} }\right) \ ,
\eea
where $\kappa \equiv \sqrt{k_1^2+k_2^2+m^2}$. Substituting \eqref{ansatz} in \eqref{dirac2} and using the Dirac representation for the matrices $\gamma^\mu$, one obtains the following differential equations for the functions $h^{I}_{\vec k}$ and $h^{II}_{\vec k}$
\bea \label{eqh}
&&\dot{h}^{I}_{\vec k}-i\left(k_3+qA\right)h^{I}_{\vec k}-i \kappa h^{II}_{\vec k}=0\ ,  \label{eqh1} \\ 
&&\dot{h}^{II}_{\vec k}+i\left(k_3+qA\right)h^{II}_{\vec k}-i \kappa h^{I}_{\vec k}=0 \label{eqh2}\ .
\eea

These equations are exactly the same as those obtained in the two-dimensional case \cite{FN}, where $\kappa$ plays here the role of the mass. %This analogy will be very useful when applying the adiabatic renormalization method in the following sections. 
With the solutions of these equations we can construct the $u$-type field modes (assumed to be of positive frequency at early times)  as follows
\be \label{u}
u_{\vec k,\lambda}(x)=\frac{e^{i\vec{k}\vec{x}}}{(2\pi)^{3/2} }\left( {\begin{array}{c}
h_{\vec k}^{I}(t)\eta_{\lambda}(\vec k) \\
h_{\vec k}^{II}(t)\lambda\eta_{\lambda}(\vec k)\\	
\end{array} }\right) \ .
\ee
Similarly, one can construct the orthogonal $v$-type field modes (of negative frequency at early times)  as 
%$\psi'=\int\frac{d^3\vec{k}}{(2\pi)^{3/2}}\psi'_{\vec k}(t)e^{-i\vec k\vec x}$ and introducing it into the Dirac equation. We finally obtain that they can be expressed as}
%The negative-frequency modes can be obtained by applying the charge conjugation operator $C\psi'=-i\gamma'^{2}\psi'^{*}=i\gamma^3\gamma^2\psi'^{*}$. {\color{blue} It also changes the sign of the electric field ($A(t)\to-A(t)$), and from equations \eqref{eqh1} and \eqref{eqh2}, one can see that it is equivalent to the changes $h^{I}_{\vec k}\to h^{II}_{-\vec k}$ , $h^{II}_{\vec k}\to h^{I}_{-\vec k}$. We finally obtain} No queda claro
\be \label{u}
v_{\vec k,\lambda}(x)=
%Cu_{\vec k,\lambda}(t)=
\frac{e^{-i\vec{k}\vec{x}}}{(2\pi)^{3/2} }\left( {\begin{array}{c}
-h_{-\vec k}^{II*}(t)\eta_{-\lambda}(-\vec k) \\
-h_{-\vec k}^{I*}(t)\lambda\eta_{-\lambda}(-\vec k)\\	
\end{array} }\right) \ .
\ee
The normalization conditions for this set of spinors, $(u_{\vec k, \lambda},v_{\vec k', \lambda'})=0$, $(u_{\vec k, \lambda},u_{\vec k', \lambda'})=(v_{\vec k, \lambda},v_{\vec k', \lambda'})=\delta^{(3)}(\vec k-\vec k')\delta_{\lambda\lambda'}$, where $(\ , \ )$ is the Dirac inner product, are ensured with the normalization condition 
\be \label{norm}
|h^{I}_{\vec k}|^2+|h^{II}_{\vec k}|^2=1 \ ,
\ee
which will be preserved on time. %, because of the properties of the Dirac inner product. 
With this set of basic spinor solutions one can construct the Fourier expansion of the Dirac field operator 
\bea
\label{spinorbd}\psi'(x)=\sum_{\lambda}\int d^3\vec k \left[B_{\vec k, \lambda} u_{\vec k, \lambda}(x)+D^{\dagger}_{\vec k, \lambda} v_{\vec k, \lambda}(x)\right] \ , 
\eea
where $B_{\vec k \lambda}$ and $D_{\vec k \lambda}$ are the annihilation operators for particles and antiparticles respectively. The normalization condition \eqref{norm} guaranties  the usual anticommutation relations for these operators: $\{B_{\vec k, \lambda},B^\dagger_{\vec k', \lambda'}\}=\{D_{\vec k, \lambda},D^\dagger_{\vec k', \lambda'}\}=\delta^3(\vec k-\vec{k'})\delta_{\lambda,\lambda'}$, and all other combinations are 0. 
\subsection{Adiabatic expansion}\label{AdExp}

Armed with the above results  we can determine a consistent  adiabatic expansion  of the four dimensional Dirac field modes interacting with the prescribed electric background. Based on the two dimensional expansion given in \cite{FN}, and taking into account that the positive-frequency solution with vanishing electric field, in the representation associated to $\psi'$, is given by
\bea
h_{\vec{k}}^{I(0)}=\sqrt{\frac{\omega-k_3}{2 \omega}}e^{-i\omega t},\\
h_{\vec{k}}^{II(0)}=-\sqrt{\frac{\omega+k_3}{2 \omega}}e^{-i\omega t},
\eea
with $\omega=\sqrt{k_3^2+\kappa^2}$,
%Note that these expressions are different from the usual ones, since we are using a non-standard representation for the gamma matrices.
 we propose the following {\it ansatz} for the field modes:
\bea \label{ansatz1}
h_{\vec{k}}^{I}=\sqrt{\frac{\omega-k_3}{2 \omega}}F(t)e^{-i \int^t \Omega(t')dt'}, \hspace{1cm}
h_{\vec{k}}^{II}=-\sqrt{\frac{\omega+k_3}{ 2\omega}}G(t)e^{-i \int^t \Omega(t')dt'},
 \eea
where the complex functions $F(t)$ and $G(t)$ and the real function $\Omega(t)$ are expanded adiabatically
%\bea \label{expansion}
%\Omega&=&\omega+\omega^{(1)}+\omega^{(2)}+\omega^{(3)}+\omega^{(4)}...\nonumber \\
%F(t)&=&1+F^{(1)}+F^{(2)}+F^{(3)}+F^{(4)}...\nonumber \\
%G(t)&=&1+G^{(1)}+G^{(2)}+G^{(3)}+G^{(4)}...\label{expansion}
%\eea
\bea \label{expansion}
\Omega(t)=\sum_{n=0}^\infty\omega^{(n)}(t),\hspace{0.7cm} F(t)=\sum_{n=0}^\infty F^{(n)}(t), \hspace{0.7cm} G(t)=\sum_{n=0}^\infty G^{(n)}(t).  \label{expansion}
\eea
Here,  $\Omega^{(n)}$, $F^{(n)}$ and $G^{(n)}$ are functions of adiabatic order $n$. The adiabatic order of a given function will be determined by its dependence on the potential vector $A(t)$ and its derivatives. In order to recover at leading order the exact solution with vanishing electric field $A(t)=0$ we demand $F^{(0)}=G^{(0)}=1$ and $\omega^{(0)}=\omega$. With this condition we are %univocally 
implicitly fixing the adiabatic order of the potential vector $A(t)$ to 1, hence, $\dot A(t)$ and $A(t)^2$ will be of order 2, $\ddot A(t)$, $A(t)\dot A(t) $ and $A(t)^3 $ of order three and so on. For a detailed discussion on the adiabatic order assignment see \cite{FNP}.\\

Plugging the ansatz \eqref{ansatz1} in the mode equations \eqref{eqh1} and \eqref{eqh2} and also in the normalization condition \eqref{norm} we get a system of equations for the functions $F(t)$, $G(t)$ and $\Omega(t)$
\bea \label{eqsFG}
(\omega - k_3)(\dot F -i\Omega F-i(k_3+qA)F)+i\kappa^2 G=0\ ,\\
(\omega + k_3)(\dot G -i\Omega G+i(k_3+qA)G)+i\kappa^2 F=0\ ,\label{eqsFG1}\\
\frac{\omega-k_3}{2\omega}|F|^2+\frac{\omega+k_3}{2\omega}|G|^2=1\label{eqsFG2}\ .
 \eea
 
 In order to obtain the expressions of the adiabatic terms $\omega^{(n)}$, $F^{(n)}$ and $G^{(n)}$, we introduce the expansion \eqref{expansion} into Eqs. \eqref{eqsFG}, \eqref{eqsFG1} and \eqref{eqsFG2} and solve them recursively, order by order. Note that $G(k_3,qA)$ satisfies the same equations as $F(-k_3,-qA)$, hence we take $G(k_3,qA)=F(-k_3,-qA)$. The system can be solved algebraically by iteration and the general solution is given by
{\small{\bea
\omega^{(n)}&=& \frac{(\omega-k_3)}{2\omega}\left[\dot  F_y^{(n-1)}-\sum_{i=1}^{n-1}\omega^{(n-i)} F_x^{(i)}-qA F_x^{(n-1)}\right]+\frac{(\omega+k_3)}{2\omega}\left[\dot  G_y^{(n-1)}-\sum_{i=1}^{n-1}\omega^{(n-i)}  G_x^{(i)}+qA G_x^{(n-1)}\right] \ , \label{w1} \\
\nonumber \\
F_x^{(n)}&=& \frac{(\omega+k_3)}{4\omega^2} \left[ \dot  F_y^{(n-1)}-\sum_{i=1}^{n-1}\omega^{(n-i)} F_x^{(i)}-qA F_x^{(n-1)}-\dot  G_y^{(n-1)}+\sum_{i=1}^{n-1}\omega^{(n-i)}  G_x^{(i)}-qA G_x^{(n-1)}  \right] \nonumber \\
&\,& %\,\,\, \,\,\,\,\,\,\,\, 
-  \frac{(\omega-k_3)}{4\omega}\sum_{i=1}^{n-1}\Big(F_x^{(i)}F_x^{(n-i)}+F_y^{(i)}F_y^{(n-i)}\Big) -\frac{(\omega+k_3)}{4\omega}\sum_{i=1}^{n-1}\Big(G_x^{(i)}G_x^{(n-i)}+G_y^{(i)}G_y^{(n-i)}\Big)   \ ,\label{F1} \\
\nonumber \\
%G_x^{(n)}&=&F_x^{(n)}(-k_3,-qA)\ ,\\
F_y^{(n)}&=&G_y^{(n)}-\frac{(\omega-k_3)}{\kappa^2}\left[\dot F_x^{(n-1)}+\sum_{i=1}^{n-1} \omega^{(n-i)}F_y^{(i)} + qA F_y^{(n-1)} \right]\ , \label{Imaginary}
\eea}}
where we have parametrized $F$ and $G$ in terms of real functions as $F=F_x+iF_y$ and $G=G_x+iG_y$. %{\color{blue}\textst{Note that $G(k_3,qA)$ satisfies the same equations as $F(-k_3,-qA)$. It means that we can take $G(k_3,qA)=F(-k_3,-qA)$, and hence, the adiabatic terms $G^{(n)}(k_3,qA)$ are exactly $F^{(n)}(-k_3,-qA)$.}} 
Note that there is an ambiguity in the imaginary part \eqref{Imaginary}. However, it disappears when computing physical observables. Further discussions on this issue are given in \cite{BFNV}.  For simplicity we choose
{\small{\bea \label{ambiguity}
F_y^{(n)}=-G_y^{(n)}=-\frac{(\omega-k_3)}{2\kappa^2}\left[\dot F_x^{(n-1)}+\sum_{i=1}^{n-1} \omega^{(n-i)}F_y^{(i)} + qA F_y^{(n-1)} \right]\ .
\eea}}
With the initial conditions $F_x^{(0)}=G_x^{(0)}=1$, $F_y^{(0)}=G_y^{(0)}=0$ and $\omega^{(0)}=\omega$ and by fixing the ambiguity according to \eqref{ambiguity},  the solutions for the adiabatic functions $F^{(n)}$,  $G^{(n)}$ and  $\omega^{(n)}$ are univocally determined. In Appendix \ref{appendixC} we give the four first terms of the adiabatic expansion.
%In Apendix A we give the details of the procedure and the expansion of the functions $F$, $G$ and $\Omega$ up to adiabatic order 4.  \\
%To obtain the expressions for ?(n), F(n), and G(n), we introduce the adiabatic expansions, and solve order by order
\section{$4$d Dirac fields: adiabatic regularization/renormalization} \label{adReg}
%To illustrate how the adiabatic regularization method works in the context of four dimensional fermions interacting with an homogeneous electric background field, we analyze explicitly a case of particular interest: the electric current. This physical observable has a major importance in this contex, since {\color{red}(...)}\\
In this section we will carry out the detailed renormalization of the vacuum expectation value of the electric current $\langle j^\mu \rangle = -q \langle \bar \psi \gamma^\mu \psi \rangle$, which constitutes the most important physical quantity in the context of strong electrodynamics \cite{Pittrich-Gies}.  The only non-vanishing component of the electric current is the one parallel to the electric field. With the results of Sec. \ref{AdExp} we can obtain the formal expression of the $z$-component of the mean electric current
\bea \label{current1}
%\langle j ^3\rangle&=&-\frac{q }{4 \pi^2 }\int_0 ^\infty k_{\perp}dk_{\perp} \int_{-\infty}^\infty dk_3(|H^{I}|^2-|H^{II}|^2-|H^{III}|^2+|H^{IV}|^2)\nonumber  \\
\langle j ^3\rangle = \frac{ 2q }{ (2 \pi)^3 }\int d^3k( |h^{II}_{\vec k}|^2-|h^{I}_{\vec k}|^2)= \frac{ q }{2 \pi^2 }\int_0 ^\infty k_{\perp}dk_{\perp} \int_{-\infty}^\infty dk_3( |h^{II}_{\vec k}|^2-|h^{I}_{\vec k}|^2)\ ,
\eea
where $k_{\perp}=\sqrt{k_1^2+k_2^2}$. This expression  is UV divergent and we have to renormalize it. The current has scaling dimension $3$, meaning that the divergences could appear up to third adiabatic order, so we have to perform adiabatic subtractions until and including the third order (note that the energy-momentum tensor requires  adiabatic subtractions of order $4$) \cite{parker-toms}. Therefore, the renormalized form of the electric current is 
% the adi In the context of quantum fields living in an homogeneous backgrounds, an since we were able to obtain the adiabatic expansion of the field modes $h_{\vec k}^{I}$ and $h_{\vec{k}}^{II}$, we can apply the adiabatic regularization method (see [ref]). In this case, the current has scaling order 3, meaning that we have to perform the adiabatic subtractions until the third adiabatic order
\bea \label{current2}
\langle j ^3\rangle_{ren}= \frac{ q }{2 \pi^2 }\int_0 ^\infty k_{\perp}dk_{\perp} \int_{-\infty}^\infty dk_3( |h^{II}_{\vec k}|^2-|h^{I}_{\vec k}|^2-\langle j ^3\rangle_{\vec k}^{(0-3)})\ ,
\eea
with $\langle j ^3\rangle_{\vec k}^{(n)}=\big(|h^{II}_{\vec k}|^2-|h^{I}_{\vec k}|^2\big)^{(n)}=-\frac{\omega-k_3}{2\omega}\sum_{i=0}^{n}F^{(i)}F^{*(n-i)}+\frac{\omega+k_3}{2\omega}\sum_{i=0}^{n}G^{(i)}G^{*(n-i)}$. These subtraction terms contain all the divergences of the electric current, giving us a finite and meaningful result for  $\langle j ^3\rangle_{ren}$. The other components give a vanishing result. After computing the subtraction terms, we finally obtain
{\small{\bea
\langle j ^3\rangle_{ren}= \frac{q }{2 \pi^2 }\int_0 ^\infty k_{\perp}dk_{\perp}  \int_{-\infty}^\infty dk_3\Biggl[( |h^{II}_{\vec k}|^2-|h^{I}_{\vec k}|^2)  -\frac{k_3}{\omega}-\frac{\kappa^2q A}{\omega^3} +\frac{3\kappa^2k_3q^2 A^2}{2\omega^5}
+\frac{(\kappa^2-4k_3^2)\kappa^2q^3 A^3}{2\omega^7}+\frac{\kappa^2q \ddot{A}}{4\omega^5}\Biggr]\ .\label{jfermions}
\eea}}
%{\color{red}
%We can compare this result with the one obtained for scalar fields.% in \ref{}
%\bea
%&&\langle j^{3}\rangle_{ren}^s=\frac{ q}{4 \pi^{2}} \int^{\infty}_{0} dk_{\perp} k_{\perp} \int_{-\infty}^{\infty}dk_3  \nonumber \\
%&&(k_3-q A) \left [|h_{\vec{k}}|^2-\frac{1}{\omega}-\frac{3 k_3^2 q^2 A^2}{2 \omega ^5}-\frac{k_3 q A}{\omega
%   ^3}+\frac{q^2 A^2}{2 \omega ^3}\right]+\frac{k_3^2 q \ddot A}{4 \omega ^5}-\frac{5 k_3^4 q^3 A^3}{2 \omega
%   ^7}+\frac{3 k_3^2 q^3 A^3}{2 \omega ^5} \label{jscalars}
%\eea
%In order to see the similarities and differences between \eqref{jfermions} and \eqref{jscalars} one can re-write the scalar current in the same way than the fermionic way (term by tem)
%\bea
%\langle j^{3}\rangle_{ren}^s=-\frac{ q}{4 \pi^{2}} \int^{\infty}_{0} dk_{\perp} k_{\perp} \int_{-\infty}^{\infty}dk_3 
%\bigl[(k_x-q A) |h_{\vec{k}}|^2 -\frac{k_3}{\omega}-\frac{\kappa^2q A}{\omega^3} +\frac{3\kappa^2k_3q^2 A^2}{2\omega^5} 
%+\frac{(\kappa^2-4k_3^2)\kappa^2q^3 A^3}{2\omega^7}+\frac{k_3^2q \ddot{A}}{4\omega^5}\bigr].\label{jscalars2}
%\eea
%We can see that the subtraction terms for the scalar current are very similars to the fermionic ones, except by a factor 2 in the last term.}
%Differences: change of sign of some terms (it could be because of the different convention of the electric charge $q$), the term with $ \ddot{A}$ is different. Factor 2.
\subsection{Conformal anomaly}
An important test  of any proposed renormalization method  is the necessary agreement with the conformal  anomaly. 
Here we compute the trace anomaly with our proposed extended adiabatic method. 
%From now on, we have shown how to perform the adiabatic expansion of four-dimensional fermionic fields and also how to compute fundamental physical observables with this expansion by using the adiabatic regularization method. However, it may happen that the proposed method does not work at all: a proper regularization method has to be able to reproduce quantum anomalies, since they are the core of quantum field theory.  For that reason, it would be very convenient to test if our regularization method works, trying to reproduce the trace anomaly.%can reproduce the trace anomaly. \\
The trace of the energy-momentum tensor is proportional to the mass of the field $\langle T^\mu_\mu \rangle=m\langle\bar\psi\psi\rangle$. Although the two point function has to be renormalized until the third adiabatic order, the trace of the energy momentum tensor must be regularized up to fourth order, i.e.,
\bea
\langle T^\mu_\mu \rangle_{ren}=m\big(\langle\bar\psi\psi\rangle_{ren}-\langle\bar\psi\psi\rangle^{(4)}\big) \ .
\eea

In the massless limit the first term vanishes, so the anomaly should appear in the subtractions of adiabatic order 4, that is% After renormalization, the physical value of the trace for $m=0$ should be 
 \bea
\langle T^\mu_\mu \rangle_{ren}=-\lim_{m\to 0}m\langle\bar\psi\psi\rangle^{(4)}\ .
\eea
The vacuum expectation value of the two-point function $\langle\bar\psi\psi\rangle$ is given by
 \bea
%\langle\bar\psi\psi\rangle&=&-\frac{1 }{4 \pi^2 }\int_0 ^\infty  k_{\perp}dk_{\perp}\int_{-\infty}^\infty dk_3(H^{I*}H^{III}+H^{III*}H^{I}+H^{II*}H^{IV}+H^{IV*}H^{I}) \nonumber \\
\langle\bar\psi\psi\rangle=\frac{1 }{2 \pi^2 }\int_0 ^\infty  k_{\perp}dk_{\perp}\int_{-\infty}^\infty dk_3\frac{m}{\kappa}(h^{I*}_{\vec k}h^{II}_{\vec k}+h^{II*}_{\vec k}h^{I}_{\vec k})\ .
\eea
By using the adiabatic regularization method, one can find the 4th order subtraction terms. Hence, in the massless limit we get
 \bea
\langle T^\mu_\mu \rangle_{ren}&=&\lim_{m\to 0}\frac{m^2 }{2 \pi^2 }\int_0 ^\infty  k_{\perp}dk_{\perp}\int_{-\infty}^\infty dk_3\frac{1}{\kappa}(h^{I*}_{\vec k}h^{II}_{\vec k}+h^{II*}_{\vec k}h^{I}_{\vec k})^{(4)}=-\frac{q^2\dot{A}^2}{12\pi^2}\ .
\eea
One can easily rewrite this result in a covariant way, obtaining the result
 \bea \label{traceA}
\langle T^\mu_\mu \rangle_{ren}=\frac{q^2}{24\pi^2}F_{\mu \nu}F^{\mu \nu}\ .
\eea
It fully agrees with the well-known result for the trace anomaly induced by an electromagnetic field for a Dirac field \cite{duff}.
%We recall that for the scalar field we have obtained (see [NF])
% \bea
%\langle T^\mu_\mu \rangle^s_{ren}=\frac{q^2}{98\pi^2}F_{\mu \nu}F^{\mu \nu}\equiv \frac{1}{4}\langle T^\mu_\mu \rangle_{ren}.
%\eea

\subsection{Relation with the DeWitt coefficients}
We will briefly see that the proposed adiabatic expansion for the fermionic modes agrees with the  Schwinger-DeWitt adiabatic expansion for the Feynman propagator. We have proved this for the adiabatic expansion of the two-dimensional theory  in Sec. \ref{section2}.  In the previous subsection we have implicitly  obtained the 4th adiabatic order, given by
\be
\langle\bar\psi\psi\rangle^{(4)}=-\frac{1}{16\pi^2m}\left(\frac23 q^2F_{\mu \nu}F^{\mu \nu}\right)=-\frac{\operatorname{tr} E_2}{16\pi^2m}
\ee
where $E_2$ coincides with the corresponding DeWitt coefficient at coincidence. Note that the numerical coefficient in the denominator is $(4\pi)^{d/2}$, where $d$ is the spacetime dimension. Moreover, at 6th adiabatic order we obtain
\be
\langle\bar\psi\psi\rangle^{(6)}=-\frac{1}{16\pi^2m}\left(\frac{2 q^2 \ddot A^2}{15 a^2}+\frac{2 q^2 A^{(3)} \dot A}{5 a^2}\right)
\ee
We can rewrite the above expression in a covariant form. It can be checked that it also fits with the DeWitt coefficient $E_3$
\be
\langle\bar\psi\psi\rangle^{(6)}=-\frac{\operatorname{tr} E_3}{16\pi^2m^3}
\ee
The general expression for $E_3$ is given in Appendix \ref{appendixA0}. Here only the flat space terms are relevant
\bea
E_3=-\frac{1}{360}\biggl(8 W_{\mu\nu ; \rho} W^{\mu\nu ; \rho}+2 W_{\mu\nu}^{ \,\,\,\,\,\,; \nu} W^{\mu\rho}_{ \,\,\,\,\,\,; \rho}+12 W_{\mu\nu ; \rho}^{\,\,\,\,\,\,\,\,\,\,\, \rho} W^{\mu\nu}-12 W_{\mu\nu} W^{\nu\rho} W_{\rho}^{\,\,\mu}+ \nonumber \\
+6 Q_{; \mu \,\,\nu}^{\,\,\,\,\mu\,\,\nu}+60 Q Q_{;\mu}^{\,\,\,\,\mu}+30 Q_{; \mu} Q^{; \mu}+60 Q^{3}+30 Q W_{\mu\nu} W^{\mu\nu}\biggr)
\eea
where $Q=-\frac{i}{2}qF_{\mu\nu}\underline \gamma^\mu\underline \gamma^\nu$ and $W_{\mu\nu}=-i q F_{\mu\nu} I$. We reinforce that the adiabatic order assignment $1$ for  $A_\mu$ is a basic ingredient for achieving the above equivalence.

\subsection{Introduction of a mass scale and renormalization ambiguities% the running of the coupling constant
}
A crucial point in the adiabatic regularization method is to fix the leading order of the adiabatic expansion, namely $\omega^{(0)}$. It seems very natural to define it as $\omega^{(0)}\equiv  \omega= \sqrt{ {\vec k}^2 +m^2 }$, as we did in Sec. \ref{AdExp}. However, there exist an inherent ambiguity in the method \cite{mu}. It is  possible to choose a slightly different  expression for the leading term $\omega^{(0)}\equiv  \omega_\mu= \sqrt{ {\vec k}^2 + \mu^2 }$, where $\mu$ corresponds to an arbitrary mass scale. In order to obtain the new adiabatic subtractions with this new choice of the leading order, one has to rewrite the mode equations as 
\bea \label{eqsmu}
i\partial_t h^{I}_{\vec k}=-(k_3+qA(t))h^{I}_{\vec k}-(\kappa_\mu+\sigma) h^{II}_{\vec k} \nonumber \\ 
i\partial_t h^{II}_{\vec k}=(k_3+qA(t))h^{II}_{\vec k}-(\kappa_\mu+\sigma) h^{I}_{\vec k}\ , 
\eea
where $\sigma=\kappa-\kappa_\mu \equiv  \sqrt{k_1^2+k_2^2+m^2} - \sqrt{k_1^2+k_2^2+\mu^2}$ is assumed of adiabatic order 1. Note that we recover the original adiabatic subtraction method by choosing $\mu=m$, and hence $\sigma=0$.\\

In this context, the ansatz of the adiabatic expansion will take the form 
\bea
h^{I}_{\vec{k}}=\sqrt{\frac{\omega_\mu-k_3}{2 \omega_\mu}}F_\mu(t)e^{-i \int^t \Omega_\mu(t')dt'}, \hspace{1cm}
h^{II}_{\vec{k}}=-\sqrt{\frac{\omega_\mu+k_3}{ 2\omega_\mu}}G_\mu(t)e^{-i \int^t \Omega_\mu(t')dt'}\label{fermMu}
\ , \eea
where  the functions $F_\mu(t)$, $G_\mu(t)$ and $\Omega_\mu(t)$ are expanded adiabatically as in \eqref{expansion}. In order to recover at order 0 the limit of vanishing electric field (and also the limit $\sigma \to 0$, since $\sigma$ is now assumed of adiabatic order 1) we demand as initial conditions $F_\mu^{(0)}=1$, $G_\mu^{(0)}=1$ and $\omega_\mu^{(0)}=\omega_\mu$.
With this new choice we can obtain the expressions of the adiabatic terms  $\omega_\mu^{(n)}$, $F_\mu^{(n)}$ and $G_\mu^{(n)}$ as before: introducing the ansatz \eqref{fermMu} in the mode equations \eqref{eqsmu} and in the normalization condition \eqref{norm}, expanding the functions $F_\mu(t)$, $G_\mu(t)$ and $\Omega_\mu(t)$ adiabatically, and finally, solving them recursively, order by order. In Appendix \ref{appendixD} we give the details of the computation and also the expression of the adiabatic renormalization subtractions for the  electric current.  
We remark that the introduction of a mass scale $\mu$ causes an unavoidable ambiguity in the renormalization procedure: it allows us to perform different adiabatic subtractions to render finite the physical observables, depending on the  scale $\mu$ we choose. For instance, concerning the renormalized current  $\langle \bar \psi \gamma^\nu \psi \rangle$  one can compare it at two different scales. Using the results given in the Appendix \ref{appendixD} we easily obtain 
\bea
\langle \bar \psi \gamma^\nu \psi \rangle_{ren} (\mu) - \langle \bar \psi \gamma^\nu \psi \rangle_{ren} (\mu_0)= \frac{q}{12\pi^2}\ln\Big(\frac{\mu^2}{\mu_0^2}\Big) \nabla_\sigma  F^{\sigma \nu} \ .  \label{currentmu}\eea

This ambiguity can be absorbed in the renormalization of the coupling constant. 
%For that reason, it is very convenient to compare the renormalized current $\langle \bar \psi \gamma^\nu \psi \rangle$ at two different scales. 
To this end it is convenient to scale the field as $\tilde A^\nu \equiv q A^\nu$ and rewrite the semiclassical Maxwell equations as
\bea
\frac{1}{q^2}\nabla_{\alpha}\tilde F^{\alpha\beta}=-\langle \bar \psi \gamma^\nu \psi \rangle_{ren} \  . \label{Maxwell1}
\eea
The above relation  for the current (\ref{currentmu}), reexpressed in terms of  $\tilde F^{\alpha\beta}$, translates into the well-known shift: $q^{-2}(\mu)-q^{-2}(\mu_0)= -(12\pi^2)^{-1}\ln\frac{\mu^2}{\mu_0^2}$, obtained within perturbative  QED using minimal subtraction in dimensional regularization \cite{qftbook}. The renormalized current given in (\ref{jfermions}) should be understood as defined at the natural scale of the problem, defined by the physical mass of the charged field, i.e., $\mu=m$ and hence $q\equiv q(m)$.

\section{Physical application: the Sauter electric pulse} \label{PhysicalApplication}
%\subsection{Schwinger limit}

As mentioned in the Introduction, one of the main advantages of the adiabatic renormalization method is its proficiency  to perform numerical computations and analytical approximations. We will devote this section to study the properties of the renormalized expression of the current  \eqref{jfermions} for the case of a pulsed electric field in a $1+3$ dimensional setting. \\
%It is interesting to study the application of the renormalized expression of the current \eqref{jfermions} for the case of a pulsed electric field. 

\begin{figure}[htbp]
\begin{center}
\begin{tabular}{c}
\includegraphics[width=90mm]{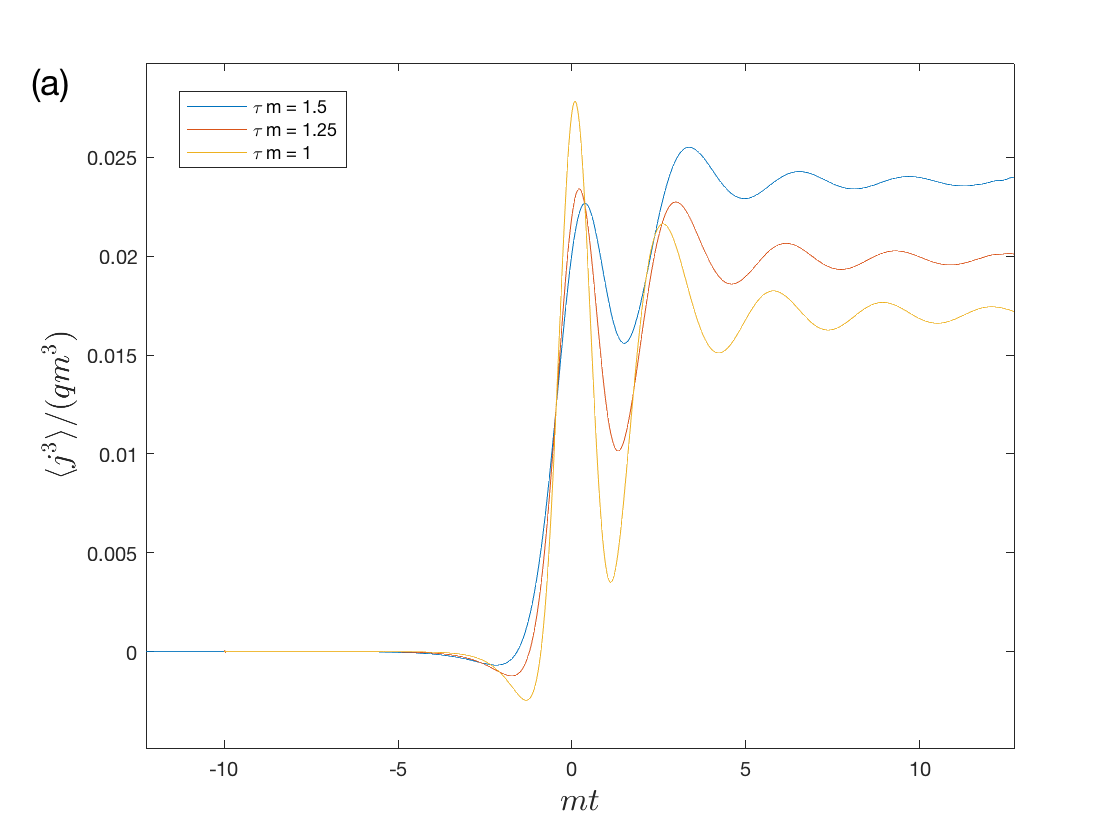}
\includegraphics[width=90mm]{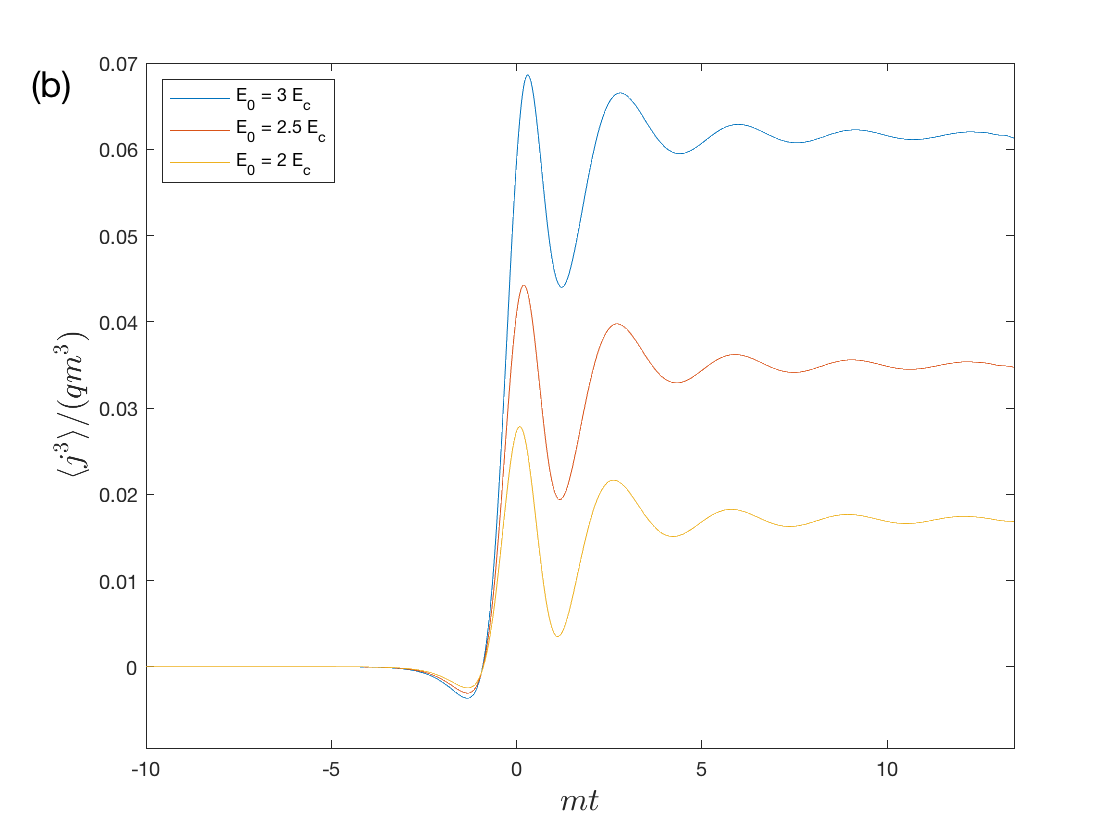}
\end{tabular}
\end{center}
\caption{\small{Evolution of the renormalized current induced by a Sauter-type electric pulse for different values of the parameters. In figure (a) the field strength  is fixed ($E_0=2E_c$), where $E_c=m^2/q$ is the critical electric field (or Schwinger limit), that is the scale above which the electric field can produce particles. In figure (b) the width of the pulse is fixed ($\tau=1/m$). We have used dimensionless variables, in terms of the mass and the charge.}}
\label{current_pulse}
\end{figure}

Let us consider the well-known Sauter-type pulse $E(t)=E_0\cosh^{-2}{(t/\tau)}$ with $\tau>0$, and its corresponding potential $A(t)=-E_0\tau\tanh{(t/\tau)}$, which is bounded at early and late times, $A(\pm\infty)=\mp E_0\tau$. This kind of pulse produces a number of particles, and then also a current, which tends to be constant when $t\to\infty$. In Fig.\ref{current_pulse} we represent the evolution of the current induced by this pulse for different values of $E_0$ and $\tau$. These figures have been obtained by solving numerically the differential equations for the modes and integrating the expression of the renormalized current \eqref{jfermions}.\\

\subsection{Late times behavior of the electric current}
We can obtain an expression of the current at late times for an electric background that vanishes at early and late times. Let us consider a pulse such that in the early and late time limits the potential is bounded as $A(-\infty)=-A_0$, $A(\infty)=A_0$, and its derivatives vanish. From Eqs. \eqref{eqh1} and \eqref{eqh2}, one can see that at late times $t \to +\infty$ the modes behave as \cite{BFNP}
\bea \label{h1inf}
h_{\vec k}^{I/II}(t)\sim  \pm \sqrt{\frac{\omega_{out} \mp (k_3+qA_0)}{2 \omega_{out}}}\alpha_{\vec k}e^{- i \omega_{out} t} + \sqrt{\frac{\omega_{out} \pm (k_3+qA_0)}{2 \omega_{out}}}\beta_{\vec k}e^{i \omega_{out} t} \ ,
\eea
where $\omega_{in/out}=\sqrt{(k_3\mp qA_0)^2 + \kappa^2}$, and $\alpha_{\vec k}$ and $\beta_{\vec k}$ are the usual Bogoliubov coefficients  satisfying  the relation $|\alpha_{\vec k}|^2+|\beta_{\vec k}|^2=1$, that ensures the normalization condition \eqref{norm}. The coefficient $|\beta_{\vec k}|^2$ gives the density number of created particles at any value of $\vec k$.\\%, and it depends on the specific form of the pulse.\\

The renormalized electric current at late times induced by an electric pulse in terms of the coefficient $|\beta_{\vec k}|^2$ can be obtained by introducing the expression of the modes at late times \eqref{h1inf} in the expression of the current \eqref{jfermions}. We obtain, for large $t$, 
\bea
\langle j ^3\rangle_{ren}&\sim& -\frac{q }{\pi^2 }\int_0 ^\infty k_{\perp}dk_{\perp}  \int_{-\infty}^\infty dk_3 \frac{k_3+qA_0}{\omega_{out}}|\beta_{\vec k}|^2 + \nonumber \\
&+& \frac{q }{2\pi^2 }\int_0 ^\infty k_{\perp}dk_{\perp}  \int_{-\infty}^\infty dk_3 \Biggl[\frac{k_3+qA_0}{\omega_{out}} -\frac{k_3}{\omega}-\frac{\kappa^2q A_0}{\omega^3} +\frac{3\kappa^2k_3q^2 A_0^2}{2\omega^5} +\frac{(\kappa^2-4k_3^2)\kappa^2q^3 A_0^3}{2\omega^7}\Biggr]\ .\label{jlate}
\eea
In Appendix \ref{appendixE} we prove that the second integral of this expression vanishes, so the current at late times is given by the simple expression
\be
\langle j ^3\rangle_{ren}\sim -\frac{q }{\pi^2 }\int_0 ^\infty k_{\perp}dk_{\perp}  \int_{-\infty}^\infty dk_3 \frac{k_3+qA_0}{\omega_{out}}|\beta_{\vec k}|^2\ .\label{jlate2}
\ee
As expected, the final current is related to the number density of particles. The analytic expression of  $|\beta_{\vec k}|^2$ depends on the form of the background.
%In order to study more in detail the behavior of the current, let us consider the well-known Sauter-type pulse $E(t)=E_0\cosh^{-2}{(t/\tau)}$ with $E_0>0$ and $\tau>0$, and its corresponding potential $A(t)=-E_0\tau\tanh{(t/\tau)}$, for which $A_0=-E_0\tau$. Solving the equation for the modes with this pulse and comparing it with \eqref{h1inf}, one can obtain the expression of the number density \cite{BFNP}

\subsection{Scaling behavior for large field strength}

It is interesting to study the behavior of the current in the limit of large field strength. To this end, we consider again the example of the Sauter pulse,  for which the coefficient $|\beta_{\vec k}|^2$ is given by (see \cite{BFNP} for more details)
 \bea
 |\beta_{\vec k}|^2=\frac{\cosh{(2\pi qE_0\tau^2)}-\cosh{(\pi(\omega_{out}-\omega_{in})\tau)}}{2\sinh{(\pi\omega_{in}\tau)}\sinh{(\pi\omega_{out}\tau)}} \label{betaqedf}\ .
 \eea 
%It is interesting to study the limit $\tau\to\infty$, which is equivalent to a constant electric field $E(t)=E_0$ to recover the Schwinger result.
Plugging it into \eqref{jlate2} we can obtain the current at late times induced by the pulse. As a test, one can compare the results given by \eqref{jlate2} with the ones given by the exact expression \eqref{jfermions} for large $t$, which are represented in Fig.\ref{current_pulse}. \\

For this pulse, assuming $qE_0>0$, the large field strength limit corresponds to $qE_0 \gg 0$.  A numerical analysis of the expression \eqref{betaqedf} shows that the relevant values of $\kappa$ and $k_3$ are of the order of $\sqrt{qE_0}$ and  $qE_0\tau$, respectively. Therefore, in order to study properly the limit of large $E_0$, it is convenient to introduce the following set of dimensionless variables
%Let us introduce, for convenience, the following set of dimensionless variables
\be \label{adimensional}
\tilde k_3=\frac{k_3}{qE_0\tau}\,\,,\,\,\tilde\kappa=\frac{\kappa}{\sqrt{qE_0}}\,\,,\,\,\ x=qE_0\tau^2 \ ,
\ee
%This choosing corresponds to the supposition that the relevant values of $k_3$ are of the order of $qE_0\tau$, and the ones of $\kappa$ are of the order of $\sqrt{qE_0}$, and therefore the adimensional parameters do not vanish in the limit of grate $E_0$. If this supposition is true, we will obtain that the... suppose that for the relevant values of $\vec k$ this adimensional parametres does not diverge
%{\color{green}and let us assume that the relevant values of $\vec k$ are such that these variables do not vanish in the limit $x\to\infty$ (we will check it later). Then the number density depends only on the parameter $x$}
and study the limit $x\to\infty$ maintaining $\tilde k_3$ and $\tilde\kappa$ constant. 
%{\color{blue} We want to study the behaviour $x \to \infty$ for fixed values of $\tilde k$ and $\tilde\kappa$ ????.
Then, we rewrite $|\beta_{\vec k}|^2$ as 
 \bea
 |\beta_{\vec k}|^2=\frac{\cosh{(2\pi x)}-\cosh{(\pi(\tilde\omega_{out}(x)-\tilde\omega_{in}(x)))}}{2\sinh{(\pi\tilde\omega_{out}(x))}\sinh{(\pi\tilde\omega_{in}(x))}} \ ,
 \eea
where $\tilde\omega_{in/out}(x)=\sqrt{x^2(\tilde k_3\pm1)^2+x\tilde\kappa^2}$. In the limit $x\to\infty$ the above expression for  $|\beta_{\vec k}|^2$ is independent of $x$, and it is given by
% \bea
% |\beta_{k}|^2\sim e^{\pi(2x-\tilde\omega_{in}(x)-\tilde\omega_{out}(x))} \ ,
% \eea
% and expanding the argument of the exponential around $\frac{\tilde\kappa}{\sqrt{x}}=0$ we get
 \bea \label{limitbeta}
 |\beta_{\vec k}|^2\sim e^{-\pi\frac{\tilde\kappa^2}{1-\tilde k_3^2}}\Theta(1-|\tilde k_3|)=e^{-\pi\frac{k_\perp^2+m^2}{qE_0}\left(\frac{1}{1-(\frac{k_3}{qE_0\tau})^2}\right)}\Theta(qE_0\tau-|k_3|) \ .
 \eea
%This means that the total number density of created particles, which is the integral in all momenta of the coefficient $|\beta_{k}|^2$, tends to a constant in the limit $qE_0\tau^2\to\infty$. In the limit $E_0>>0$.
%Note that if we had chosen different set of dimensionless variables we would have obtained an apparent different behavior. For example for $\tilde\kappa=\frac{\kappa}{qE_0\tau}$ one obtains $ |\beta_{\vec k}|^2\to0$ for large $x$, which has no sense since a large electric field must produce a large amount of particles.\\

%The result (\ref{limitbeta}) tells us that the relevant values of $\kappa$ and $k_3$ are of the order of $\sqrt{qE_0}$ and  $qE_0\tau$, respectively.
%which justifies, a posteriori, the choice  of the variables \eqref{adimensional}.}
Substituting the expression \eqref{limitbeta} into \eqref{jlate2} and taking into account that $\frac{k_3-qE_0\tau}{\omega_{out}}\sim-1$ for large $E_0$, we obtain the behavior  of  the current at late times created by a high intensity  pulse
%\be \label{jlargeE0}
%\langle j ^3\rangle_{ren}^{t\to\infty}\sim \frac{q^2E_0}{2\pi^3}  \int_{-qE_0\tau}^{qE_0\tau} dk_3 \left(1-(\frac{k_3}{qE_0\tau})^2\right) e^{-\pi\frac{m^2}{qE_0}\left(\frac{1}{1-(\frac{k_3}{qE_0\tau})^2}\right)}\,.
%\ee
\be \label{jlargeE0}
\langle j ^3\rangle_{ren}\sim \frac{q^3E_0^2\tau}{2\pi^3}  \int_{-1}^{1} ds (1-s^2) e^{-\pi\frac{m^2}{qE_0}\left(\frac{1}{1-s^2}\right)}\ .
\ee

%This integral has not analytic solution, but it can be solved numerically.
 Assuming now that $qE_0>>m^2$, the above integral \eqref{jlargeE0} can be done exactly and we finally obtain
%Taking into account that $E_0>>m$, we finally obtain the expression of the current for large intensisty of the pulse
\be \label{StrongField}
\langle j ^3\rangle_{ren}\sim\frac{2}{3\pi^3}q^3E_0^2\tau\ ,
\ee
%The result \eqref{NSch} is valid for the approximation of constant electric field, but in order to obtain the exact result for the Sauter pulse in the limit $E_0\tau^2>>0$ we must consider the expression \eqref{limitbeta}, which depends on $k_3$.
which is the predicted expression of the current in the limit of large field strength $E_0$. We can also obtain the total number density of created quanta for the Sauter pulse in this limit
\bea \label{NSauter}
\langle N\rangle=\sum_\lambda\int\frac{d^3k}{(2\pi)^3}(|\beta_{\vec k}|^2+|\beta_{-\vec k}|^2)\sim\frac{2}{3\pi^3}q^2E_0^2\tau%e^{-\pi\frac{m^2}{qE_0}}
\ .
\eea

It is interesting to compare the result \eqref{StrongField} with the one obtained for a scalar field. The coefficient $|\beta_{\vec k}|^2$ in this case has a different expression, but it tends to the same limit for large $E_0$ \eqref{limitbeta}. Therefore the scaling behavior of the current at late times ($\langle j ^3\rangle_{ren}^{scalar} \sim\frac{1}{3\pi^3}q^3E_0^2\tau$) will be the same as in the fermionic case, except for the factor $2$, on account of the absence of the spin degree of freedom.  
\\
%\subsubsection{Slowly varying pulse}

For completeness, it is worth  to see how the above results  can also serve to describe the  Schwinger limit, i.e, a constant electric field. Note that the expression \eqref{limitbeta} has been obtained for the limit $E_0\tau^2>>0$, so it should also be valid for the limit of large $\tau$, keeping $E_0$ constant, which describes a pulse with a large width. Bringing this limit to the extreme case $\tau\to\infty$, we get $|\beta_{\vec k}|^2\sim \exp{(-\pi\frac{k_\perp^2+m^2}{qE_0})}$, which is the well-known expression for the beta coefficients of a constant electric field \cite{narozhnyi1970} leading to the Schwinger formula for the vacuum persistence amplitude.

\section{Conclusions}{\label{conclusions}
In this work we have extended %implemented a satisfactory extension of 
the adiabatic regularization method for 4-dimensional Dirac fields interacting with a time-varying electric background. Our approach can be distinguished from previous analysis in the literature in the adiabatic order assignment for the vector potential, which is chosen to be of order $1$. This choice is required to fit it with the expected equivalence with the Schwinger-DeWitt adiabatic expansion.  Our proposal has required to introduce a nontrivial ansatz, Eq. \eqref{ansatz1}, to generate a self-consistent adiabatic expansion of the fermionic modes. The given expansion turns out to be different from the    
  WKB-type expansion used for scalar fields. With this extension we have obtained a well-defined prediction, Eq. \eqref{jfermions}, for the renormalized electric current induced by the created particles.
Our proposal is consistent, in the massless limit, with the conformal anomaly. The expected equivalence with the Schwinger-DeWitt expansion is explicitly realized. % and we have also accounted for the effective scaling of the electric coupling constant. 
%We have tested the consistency of the method by reproducing the well known trace anomaly \eqref{traceA}. 
In parallel we have also explored the physical consequences of the introduction of an arbitrary mass scale on the adiabatic regularization scheme, finding consistency with the behavior of the effective scaling of the electric coupling constant.  To illustrate the power of the method we have analyzed the pair production phenomenon in the particular case of a Sauter-type electric pulse. %{\color{blue}\textst{We have given a well-defined prediction for the renormalized electric current induced by the electric pulse.}} 
In particular, we have obtained the scaling behavior of the  current in the strong field regime [Eq. \eqref{StrongField}].

%We have tested the consistency of the method by reproducing the well known trace anomaly \eqref{traceA}. In parallel we have also explored the physical consequences of the introduction of an arbitrary mass scale on the adiabatic regularization scheme, finding consistency with the results of the trace anomaly \eqref{rq}. Finally, we have analyzed the particular case of an electric pulse. We have studied the particle production phenomenon in terms of the electric current \eqref{jfermions}. In particular, we have obtained the electric current in the strong field regime \eqref{StrongField}. Our results are consistent with the Schwinger limit.
\section*{Acknowledgments}
 We thank I. Agullo, P. R. Anderson,  A. del Rio and A. Ferreiro for very useful comments. This work was supported by the Spanish MINECO research Grants No. FIS2017-84440-C2-1-P and No. FIS2017-91161- EXP, and by the Generalitat Valenciana Grant No. SEJI/2017/042. P. B. is supported by the Formaci\'on del Personal Universitario Ph.D. fellowship FPU17/03712. S. P.  is supported by the Formaci\'on del Personal Universitario Ph.D. fellowship FPU16/05287.
\appendix
 \section{DEWITT COEFFICIENT $E_3$} \label{appendixA0}

The  expression for the DeWitt coefficient of sixth adiabatic order is \cite{Gilkey, Vassilevich}
\bea
E_3&=&-\frac{1}{7 !}\biggl(-18 R_{; \mu \,\,\nu}^{\,\,\,\,\mu\,\,\nu}+17 R_{; \mu} R^{; \mu}-2 R_{\mu \nu ; \rho} R^{\mu \nu ; \rho}-4 R_{\mu\nu ; \rho} R^{\mu \rho ; \nu}+9 R_{\mu\nu\rho\sigma;\alpha} R^{\mu\nu\rho\sigma;\alpha}+28 R R_{;\mu}^{\,\,\,\,\mu}-8 R_{\mu\nu} R^{\mu\nu\,\,\,\,\rho}_{\,\,\,\,\,\,;\rho}+ \nonumber \\
&+&24 R_{\mu\nu} R^{\mu \rho; \nu}_{ \,\,\,\,\,\,\,\,\,\,\,\rho}+12 R_{\mu\nu\rho\sigma} R^{\mu\nu\rho\sigma;\alpha}_{\,\,\,\,\,\,\,\,\,\,\,\,\,\,\,\,\,\,\alpha}-\frac{35}{9} R^{3}+\frac{14}{3} R R_{\mu\nu} R^{\mu\nu} -\frac{14}{3} R R_{\mu\nu\rho\sigma} R^{\mu\nu\rho\sigma}+\frac{208}{9} R_{\mu\nu} R^{\mu\rho} R^{\nu}_{\,\,\rho}- \nonumber \\
&-&\frac{64}{3} R_{\mu\nu} R_{\rho\sigma} R^{\mu\rho\nu\sigma}+\frac{16}{3} R_{\mu}^{\,\,\,\nu} R^{\mu\rho\sigma\alpha} R_{\nu\rho\sigma\alpha}-\frac{44}{9} R_{\mu\nu\rho\sigma} R^{\mu\nu\alpha\beta} R^{\rho\sigma}_{\,\,\,\,\,\,\alpha\beta}-\frac{80}{9} R_{\mu\nu\rho\sigma} R^{\mu\alpha\rho\beta} R^{\nu\,\,\sigma}_{\,\,\alpha\,\,\beta}\biggr)I- \nonumber \\
&-&\frac{1}{360}\biggl(8 W_{\mu\nu ; \rho} W^{\mu\nu ; \rho}+2 W_{\mu\nu}^{ \,\,\,\,\,\,; \nu} W^{\mu\rho}_{ \,\,\,\,\,\,; \rho}+12 W_{\mu\nu ; \rho}^{\,\,\,\,\,\,\,\,\,\,\, \rho} W^{\mu\nu}-12 W_{\mu\nu} W^{\nu\rho} W_{\rho}^{\,\,\mu}-6 R_{\mu\nu\rho\sigma} W^{\mu\nu} W^{\rho\sigma}+4 R_{\mu\nu} W^{\mu\rho} W^{\nu}_{\,\,\rho}- \nonumber \\
&-&5 R W_{\mu\nu} W^{\mu\nu}+6 Q_{; \mu \,\,\nu}^{\,\,\,\,\mu\,\,\nu}+60 Q Q_{;\mu}^{\,\,\,\,\mu}+30 Q_{; \mu} Q^{; \mu}+60 Q^{3}+30 Q W_{\mu\nu} W^{\mu\nu}-10 R Q_{;\mu}^{\,\,\,\,\mu}-4 R_{\mu\nu} Q^{;\mu\nu}-12 R_{; \mu} Q^{; \mu}- \nonumber \\
&-&30 Q^2 R-12 Q R_{;\mu}^{\,\,\,\,\mu}+5 Q R^{2}-2 Q R_{\mu\nu} R^{\mu\nu}+2 Q R_{\mu\nu\rho\sigma} R^{\mu\nu\rho\sigma}\biggr)
\eea
where, for scalar fields $Q=\xi R$, $W_{\mu\nu}=i q F_{\mu\nu}$ and $I=1$, while for Dirac fields $Q=\frac{1}{4}R I-\frac{i}{2}qF_{\mu\nu}\underline \gamma^\mu\underline \gamma^\nu$, $W_{\mu\nu}=-i q F_{\mu\nu} I-\frac14 R_{\mu\nu\rho\sigma} \underline \gamma^\rho\underline\gamma^\sigma$ and $I$ is the identity matrix.

\section{MATCHING HADAMARD COEFFICIENTS WITH SCALAR ADIABATIC REGULARIZATION} \label{appendixB}
In this appendix, we relate the adiabatic regularization method with Hadamard renormalization for charged scalar fields. To simplify the comparison we will restrict the analysis to Minkowski spacetime. In Sec. \ref{section2} we have introduced the basics of the adiabatic regularization method for 4-dimensional charged scalar fields interacting with an electromagnetic background. The renormalized vacuum expectation value on the two point function was given in \eqref{phiphiA}. For the  electric current, defined as $j_\mu=iq[\phi^{\dagger}D^{\mu}\phi-(D^{\mu}\phi)^{\dagger}\phi]$, we obtain
%\bea
%\langle j^{3}\rangle_{ren}&=&\frac{ q}{(2 \pi)^{3}} \int d^3k
%\Big[k_3 |h_{\vec{k}}|^2-\Big(k_3 |h_{\vec{k}}|^2\Big)^{(0-3)}\Big]-qA\langle \phi \phi^\dagger  \rangle_{ren}\label{jscalars}
%\eea
\bea
\langle j^{3}\rangle_{ren}&=&\frac{ q}{(2 \pi)^{3}} \int d^3k
\Big[(k_3-qA) |h_{\vec{k}}|^2 -\langle j^3 \rangle_{\vec{k}}^{(0-3)}\Big]\ ,\label{jscalars}
\eea
with $\langle j^3 \rangle_{\vec{k}}^{(0-3)}=\sum_{n=0}^3k_3(\Omega_{\vec{k}}^{-1})^{(n)}-qA\langle \phi^\dagger  \phi \rangle_{\vec{k}}^{(0-2)}$. To compute these subtraction terms, we usually fix the leading order of the adiabatic expansion as $\omega^{(0)}=\omega=\sqrt{\vec{k}^2+m^2}$.\\

As explained in the main text, the choice of the leading term is crucial to define the adiabatic expansion. We have argued that to properly fix the leading therm we have to choose the vector potential $A$ of adiabatic order 1. %  To this end, we firstly remark that
However, the choice of the leading term $\omega^{(0)}$ is not yet completely fixed, and one can make a more general choice defining $\omega^{(0)}=\omega_\mu=\sqrt{{\vec{k}}^2+\mu^2}$, where $\mu$ is an arbitrary mass scale.  With this new choice the adiabatic expansion can be re-calculated, giving us slightly different subtraction terms. An exhaustive analysis of this ambiguity can be found in \cite{mu}.
%This ambiguity gives us different possibilities of making the adiabatic subtractions, depending on the scale parameter $\mu $ we choose. 
The ambiguity on the subtractions, leads to an ambiguity on the physical observables. For the two point function the ambiguity manifests as
\bea \label{twopointAd1}
\langle \phi^{\dagger} \phi \rangle_{ren}(\mu)=\langle \phi^{\dagger} \phi\rangle_{ren}(\mu_0)-\frac{\alpha}{2}\Big[m^2\ln\Big(\frac{\mu^2}{\mu_0^2}\Big)-\mu^2+\mu_0^2\Big]\ ,
\eea
where $\alpha=\frac{1}{2 (2\pi)^2}$, and for the electric current we find
\bea
\langle j^3\rangle_{ren}(\mu)=\langle j^3\rangle_{ren}(\mu_0)-\frac{\alpha}{6}\ln\Big(\frac{\mu^2}{\mu_0^2}\Big)q^2 \ddot A\ .
\eea
Rewriting the equation above in a covariant way, we get
\bea \label{currentAd1}
\langle j^\nu\rangle_{ren}(\mu)=\langle j^\nu\rangle_{ren}(\mu_0)-\frac{q^2\alpha}{6}\ln\Big(\frac{\mu^2}{\mu_0^2}\Big) \nabla_\sigma F^{\sigma \nu}\ .
\eea

\subsubsection*{ Matching with Hadamard renormalization}

We can compare the results summarized in Sec. II  with the results given by Hadamard renormalization, particularizing for the case in which $A_\mu=(0,0,0,-A(t))$. Adopting the notation given in  \cite{Had1}, the expectation value of the two point function can be expressed as
%Valores de $\alpha$\\
% $\alpha=\frac{1}{4\pi}$ for $d=2$ and 
 %$\alpha=\frac{1}{2 (2\pi)^2}$ for $d=4$\\
% Asumimos $A_\mu=(0,0,0,-A)$ pero hay que revisar el signo, ellos usan convenciones distintas a las nuestras.\\
\bea
\langle \phi \phi^\dagger  \rangle _{ren}=\alpha w_0(x)\ ,\label{phiphiH}
\eea
and the electric current is given by
%\bea
%\langle J_\mu \rangle_{ren} =\frac {\alpha q }{4 \pi} (qA_\mu w_0(x)-\Im[w_{1\mu}(x)])
%\eea
%Tercera componente
%\bea
%\langle J_3\rangle_{ren} =\frac {\alpha q }{4 \pi} ( -q A w_0(x)-\Im[w_{13}(x)])\label{j3H}
%\eea
%Nota: para ellos $J_\mu=\frac{iq}{8\pi}(\phi^* D_\mu\phi-\phi (D_\mu\phi)^* )$ mientras que para nosotros  $j_\mu=iq(\phi^* D_\mu\phi-\phi (D_\mu\phi)^* )$. Por tanto 
\bea\langle j_\mu \rangle=-2 q\alpha  (qA_\mu w_0(x)+\Im[w_{1\mu}(x)])\label{jH}\ ,\eea
where $\alpha=\frac{1}{2 (2\pi)^2}$ and the functions $w_0$ and $w_{1\mu}$ are the first terms of the covariant Taylor series expansion of the Hadamard biscalar $W(x,x')$. \\

Comparing \eqref{phiphiH} with \eqref{phiphiA} we immediately get
\bea
\alpha w_0=\frac{ 1}{2(2 \pi)^{3}} \int d^3k\left[|h_{\vec{k}}|^2-\sum_{n=0}^2(\Omega_{\vec{k}}^{-1})^{(n)}\right]\ , 
\eea
and hence, by using the previous result and Eqs. \eqref{jscalars} and \eqref{jH} we directly find
\bea
\alpha \Im (w_{13})=\frac{ q}{2(2 \pi)^{3}} \int d^3k
\Big[k_3 |h_{\vec{k}}|^2-\sum_{n=0}^3k_3(\Omega_{\vec{k}}^{-1})^{(n)}\Big]\ .
\eea
Hadamard renormalization scheme also presents a renormalization ambiguity in even space-time dimensions, due to the choice of the renormalization lenght scale $\ell$. %In this case, the ambiguity is manifested in the Hadamard coefficients as
%
%\bea
%w_{0} & \rightarrow & w_{0}+V_{00} \ln \ell^{2}\\ w_{1 \mu}  & \rightarrow & w_{1 \mu}+V_{01 \mu} \ln \ell^{2}
%\eea
%\bea
%\langle J_{\mu}\rangle_{\mathrm{ren}} \rightarrow \frac{\alpha q}{4 \pi}\left\{q A_{\mu} w_{0}-\Im\left[w_{1 \mu}\right]+\frac{q}{12}\left(\nabla^{\rho} F_{\rho \mu}\right) \ln \ell^{2}\right\}
%\eea
The ambiguity is manifested in the physical observables as 
\bea
&&\langle \phi \phi^\dagger  \rangle _{ren} \rightarrow \langle \phi \phi^\dagger  \rangle _{ren}+\frac{\alpha}{2}m^2 \ln \ell^2\ ,\\
&&\langle j_{\mu}\rangle_{ren} \rightarrow\langle j_{\mu}\rangle_{ren}+ \frac{\alpha q^2}{6}\left(\nabla^{\rho} F_{\rho \mu}\right) \ln \ell^{2}\ .
\eea
Note that the length scale $\ell$ is inversely proportional to the mass scale $\mu$. Comparing these results with the ones obtained with adiabatic regularization [Eqs. \eqref{twopointAd1} and \eqref{currentAd1}] we find that the logarithmic part of the ambiguity is exactly the same. However, with adiabatic regularization we also find a quadratic term in the ambiguity of the two point function. %, which goes as $\mu^2$.

\section{SUBTRACTION TERMS } \label{appendixC}
In this appendix we give the explicit expressions of the adiabatic expansion of the fermionic field modes up to and including the fourth adiabatic order. We remind that $G^{(n)}(k_3,qA)=F^{(n)}(-k_3,-qA)$.\\

 \textit{Order 0}
 \bea
 \omega^{(0)}&=&\omega, \hspace{0.5cm}F_x^{(0)}=G_x^{(0)}=1, \hspace{0.5cm}F_y^{(0)}= G_y^{(0)}=0\ .
 \eea
 
 \textit{Order 1}
 \bea
 \omega^{(1)}=\frac{q A k_3}{\omega },\hspace{0.5cm}
 F_x^{(1)}=-\frac{q A \left(\omega +k_3\right)}{2 \omega ^2},\hspace{0.5cm}
 %  G_x^{(1)}&=&\frac{q A \left(\omega -k_3\right)}{2 \omega ^2}\\ 
    F_y^{(1)}= G_y^{(1)}=0\ .
 \eea
 
  \textit{Order 2}
 \bea
 \omega^{(2)}=\frac{q^2 A^2 \kappa ^2}{2 \omega ^4},\hspace{0.5cm}
   %F_x^{(2)}&=&-\frac{5 q^2 A^2 \kappa ^2}{8 \omega ^4}+\frac{q^2 A^2}{2 \omega ^2}+\frac{q^2 A^2 k_3}{2\omega ^3}\\ 
      F_x^{(2)}=-\frac{5 q^2 A^2 \kappa ^2 }{8 \omega ^4}+\frac{q^2A^2(\omega+k_3 )}{2 \omega ^3},\hspace{0.5cm}
  %    G_x^{(2)}&=&-\frac{5 q^2A^2 \kappa ^2 }{8 \omega ^4}+\frac{q^2A^2 (\omega-k_3 )}{2 \omega ^3}\\ 
    F_y^{(2)}=- G_y^{(2)}=\frac{q\dot{A} }{4 \omega ^2}\ .
      %G_y^{(2)}&=&-\frac{q\dot{A} }{4 \omega ^2}
 \eea
 %we choose $N(t)=-\frac{\dot{A} q}{4 \omega ^2}$\\
 
   \textit{Order 3}
 \bea
 \omega^{(3)}&=&-\frac{q^3 A^3 \kappa ^2 k_3}{2 \omega ^5}-\frac{q \ddot{A} k_3}{4 \omega ^3},\\
 F_x^{(3)}&=&\frac{11 q^3 A^3 \kappa ^2}{16 \omega ^5}-\frac{q^3 A^3}{2 \omega ^3}+\frac{15 q^3 A^3 \kappa ^2 k_3}{16 \omega ^6}-\frac{q^3 A^3 k_3}{2 \omega ^4}+\frac{q \ddot{A}  \left(\omega +k_3\right)}{8 \omega ^4},\\
  % G_x^{(3)}&=&%\frac{q^3A^3  (\omega-k_3 ) \left(4 \omega  (k_3+3 \omega )-15\kappa ^2\right)}{16 \omega ^6}-\frac{q \ddot{A} \left(\omega -k_3\right)}{8 \omega ^4}\\
  %  -\frac{11 q^3 A^3 \kappa ^2}{16 \omega ^5}+\frac{q^3 A^3}{2 \omega ^3}+\frac{15 q^3 A^3\kappa ^2 k_3}{16 \omega ^6}-\frac{q^3 A^3 k_3}{2 \omega ^4}-\frac{q \ddot{A} \left(\omega -k_3\right)}{8 \omega ^4}\\ 
F_y^{(3)}&=&-G_y^{(3)}=-\frac{5q^2 A \dot{A} k_3 }{8 \omega ^4}\ .
 \eea
% we choose $M(t)=\frac{5 A \dot{A} k_3 q^2}{8 \omega ^4}$\\  
  
    \textit{Order 4} 
 \bea
 \omega^{(4)}&=&%-\frac{5 A^4 \kappa ^4 q^4}{8 \omega ^7}+\frac{A^4 \kappa ^2 q^4}{2 \omega ^5}+\frac{A k_3  q^2 \ddot{A}}{8 \omega ^4}+\frac{\dot{A}^2 k_3 q^2}{8 \omega ^4}-\frac{A \kappa ^2 q^2\ddot{A}}{\omega ^5}-\frac{7 \dot{A}^2 \kappa ^2 q^2}{8 \omega ^5}+\frac{A q^2 \ddot{A}}{2 \omega ^3}+\frac{\dot{A}^2 q^2}{2 \omega ^3}
 -\frac{5 q^4 A^4 \kappa ^4 }{8 \omega ^7}+\frac{q^4A^4 \kappa ^2}{2 \omega ^5}-\frac{3\kappa ^2 q^2 A   \ddot A}{4 \omega^5}+\frac{5q^2 A  \ddot A}{8 \omega ^3}+\frac{5 k_3^2 q^2 \dot A^2}{8 \omega ^5},\\
 F_x^{(4)}&=&-\frac{17 A^4 \kappa ^2 k_3 q^4}{16 \omega ^7}+\frac{A^4 k_3 q^4}{2 \omega ^5}+\frac{195
   A^4 \kappa ^4 q^4}{128 \omega ^8}-\frac{31 A^4 \kappa ^2 q^4}{16 \omega ^6}+\frac{A^4
   q^4}{2 \omega ^4}-\frac{A k_3 q^2 \ddot{A}}{2 \omega ^5}\nonumber \\
   &-&\frac{5 \dot{A}^2 k_3 q^2}{16
   \omega ^5}+\frac{9 A \kappa ^2 q^2 \ddot{A}}{16 \omega ^6}+\frac{5 \dot{A}^2 \kappa ^2
   q^2}{16 \omega ^6}-\frac{A q^2 \ddot{A}}{2 \omega ^4}-\frac{11 \dot{A}^2 q^2}{32 \omega
   ^4}, \\
%G_x^{(4)}&=&+\frac{17 A^4 \kappa ^2 k_3 q^4}{16 \omega ^7}-\frac{A^4 k_3 q^4}{2 \omega ^5}+\frac{195 A^4
%   \kappa ^4 q^4}{128 \omega ^8}-\frac{31 A^4 \kappa ^2 q^4}{16 \omega ^6}+\frac{A^4 q^4}{2
 %  \omega ^4}+\frac{A k_3 q^2 \ddot{A}}{2 \omega ^5}\nonumber \\
  % &+&\frac{5 \dot{A}^2 k_3 q^2}{16 \omega
   %^5}+\frac{9 A \kappa ^2 q^2 \ddot{A}}{16 \omega ^6}+\frac{5 \dot{A}^2 \kappa ^2 q^2}{16
   %\omega ^6}-\frac{A q^2 \ddot{A}}{2 \omega ^4}-\frac{11 \dot{A}^2 q^2}{32 \omega ^4}\\
F_y^{(4)}&=&-G_y^{(4)}=\frac{q^3A^2 \dot A \left(34 \omega ^2-45 \kappa ^2\right)}{32
   \omega ^6}-\frac{qA^{(3)} }{16 \omega ^4}\ .
 \eea \\
\\
\section{$\mu$-PARAMETER ADIABATIC EXPANSION} \label{appendixD}
The general solution for $F_\mu^{(n)}$,  $G_\mu^{(n)}$ and $\omega_\mu^{(n)}$ is given by

{\small{\bea
%\omega_\mu^{(n)}&=&\frac{1}{2\omega_\mu}\Bigg[ (\omega_\mu-k_3)\left((\dot F_\mu)_y^{(n-1)}-\sum_{i=1}^{n-1}(\omega_\mu)^{(n-i)} (F_\mu)_x^{(i)}-qA (F_\mu)_x^{(n-1)}\right)\nonumber \\
%&\,  \,&\,\,\,\,\,\,\, \,\,\,+(\omega_\mu+k_3)\left( (\dot G_\mu)_y^{(n-1)}-\sum_{i=1}^{n-1}(\omega_\mu)^{(n-i)}  (G_\mu)_x^{(i)}+qA (G_\mu)_x^{(n-1)}\right) \nonumber \\
%&\,\,\,&\,\,\,\,\,\,\,\,\,\,+ 2\sigma \kappa_\mu \Big[ (G_\mu)_x^{(n-1)}+(F_\mu)_x^{(n-1)}\Big]+\sigma^2  \Big[ (G_\mu)_x^{(n-2)}+(F_\mu)_x^{(n-2)} \Big] \Bigg], \\
%\nonumber \\
\omega_\mu^{(n)}&=&\omega^{(n)}(\omega_\mu,\kappa_\mu,F_\mu,G_\mu)+ \frac{\sigma \kappa_\mu}{\omega_\mu} \Big[ (G_\mu)_x^{(n-1)}+(F_\mu)_x^{(n-1)}\Big]+\frac{\sigma^2}{2\omega_\mu}  \Big[ (G_\mu)_x^{(n-2)}+(F_\mu)_x^{(n-2)} \Big]\ , \\
\nonumber \\
(F_{\mu })_x^{(n)}&=&F_x^{(n)}(\omega_\mu,\kappa_\mu,F_\mu,G_\mu)+\frac{1}{4\omega_\mu^2} \Bigg\{ \frac{\omega_\mu+k_3}{\omega_{\mu}-k_3}\Big[2\sigma \kappa_{\mu} (G_\mu)_x^{(n-1)}+\sigma^2   (G_\mu)_x^{(n-2)}\Big]-2\sigma \kappa_\mu (F_\mu)_x^{(n-1)}-\sigma^2   (F_\mu)_x^{(n-2)} \Bigg\}\ ,\\
\nonumber \\
%(G_\mu)_y^{(n)}&=&G_y^{(n)}(\omega_\mu,\kappa_\mu,F_\mu,G_\mu)+\frac{1}{\kappa_\mu^2} \Big[ 2\sigma \kappa_\mu (G_\mu)_y^{(n-1)}+\sigma^2   (G_\mu)_y^{(n-2)}\Big] \label{Imaginary2bis}
(F_\mu)_y^{(n)}&=&F_y^{(n)}(\omega_\mu,\kappa_\mu,F_\mu,G_\mu)+\frac{1}{\kappa_\mu^2} \Big[ 2\sigma \kappa_\mu (G_\mu)_y^{(n-1)}+\sigma^2   (G_\mu)_y^{(n-2)}\Big]\ , \label{Imaginary2bis}
\eea}}
where $\omega^{(n)}/F^{(n)}/G^{(n)}(\omega_\mu,\kappa_\mu,F_\mu,G_\mu)$ are given by the expressions \eqref{w1}, \eqref{F1} and \eqref{Imaginary} with the changes $(\omega,\kappa,F,G)\to(\omega_\mu,\kappa_\mu,F_\mu,G_\mu)$. Note again that $G_\mu(k_3,qA)$ satisfies the same equations than $F_\mu(-k_3,-qA)$, and hence   $G_\mu^{(n)}(k_3,qA)=F_\mu^{(n)}(-k_3,-qA)$. We also find an ambiguity in the imaginary part \eqref{Imaginary2bis}. For simplicity we choose

{\footnotesize{\bea \label{ambiguity1}
(F_\mu)_y^{(n)}=-(G_\mu)_y^{(n)}=-\frac{(\omega_\mu-k_3)}{2\kappa_\mu^2}\left[(\dot F_\mu)_x^{(n-1)}+\sum_{i=1}^{n-1} \omega_\mu^{(n-i)}(F_\mu)_y^{(i)} + qA (F_\mu)_y^{(n-1)} -\frac{1}{\omega_\mu-k_3}\Big( 2\sigma \kappa_\mu (G_\mu)_y^{(n-1)}+\sigma^2   (G_\mu)_y^{(n-2)}\Big)\right]
\eea}}

With the initial conditions $(F_\mu)_x^{(0)}=(G_\mu)_x^{(0)}=1$, $(F_\mu)_y^{(0)}=(G_\mu)_y^{(0)}=0$ and $\omega_\mu^{(0)}=\omega_\mu$ and by fixing the ambiguity \eqref{ambiguity1},  the solutions for the adiabatic functions $F_\mu^{(n)}$,  $G_\mu^{(n)}$ and  $\omega_\mu^{(n)}$ are univocally determined. \\

The renormalized electric current for an arbitrary mass scale is given by
\bea
\langle j ^3\rangle_{ren}&=& \frac{q }{2 \pi^2 }\int_0 ^\infty k_{\perp}dk_{\perp}  \int_{-\infty}^\infty dk_3\Biggl[( |h^{II}_{\vec k}|^2-|h^{I}_{\vec k}|^2)  - \langle j^3\rangle_{\vec k}^{(0-3)}(\mu)\Biggr]\ ,\label{jfermionsmu}
\eea
 with
 \bea
 \langle j^3\rangle_{\vec k}^{(0)}(\mu)&=&\frac{k_3}{\omega_\mu}\ ,\\
  \langle j^3\rangle_{\vec k}^{(1)}(\mu)&=&\frac{\kappa_\mu^2q A}{\omega_\mu^3}-\frac{2k_3\kappa_\mu \sigma}{\omega_\mu^3}\ ,\\
   \langle j^3\rangle_{\vec k}^{(2)}(\mu)&=&-\frac{3\kappa^2k_3q^2 A^2}{2\omega_\mu^5}-\frac{2qA\kappa_\mu \sigma(3\kappa_\mu^2-2\omega_\mu^2)}{\omega_\mu^5}+\frac{3k_3 \sigma^2(2\kappa_\mu^2-\omega_\mu^2)}{\omega_\mu^5}\ ,\\
    \langle j^3\rangle_{\vec k}^{(3)}(\mu)&=&+\frac{\kappa_\mu^2q^3 A^3(4\omega_\mu^2-5\kappa_\mu^2)}{2\omega_\mu^7}-\frac{2k_3\sigma^3(10 \kappa_\mu^4-9\kappa_\mu^2\omega_\mu^2+\omega_\mu^4)}{\kappa\omega_\mu^7}+\frac{3k_3q^2A^2\sigma\kappa_\mu(5\kappa_\mu^2-2\omega_\mu^2)}{\omega_\mu^7}\\   &\,&\,  +\frac{3qA\sigma^2(10\kappa_\mu^4-11\kappa_\mu^2\omega_\mu^2+2\omega_\mu^4)}{\omega_\mu^7}-\frac{\kappa_\mu^2q \ddot{A}}{4\omega_\mu^5}\ .\nonumber
 \eea
 
\section{SIMPLIFICATION OF THE EXPRESSION OF THE CURRENT AT LATE TIMES} \label{appendixE}

In this appendix we prove that the second integral in the expression of the current at late times (see Eq. \eqref{jlate}), %vanishes. %The relevant integral is
\be
I=\int_0 ^\infty k_{\perp}dk_{\perp}  \int_{-\infty}^\infty dk_3 \Biggl[\frac{k_3+qA_0}{\omega_{out}} -\frac{k_3}{\omega}-\frac{\kappa^2q A_0}{\omega^3} +\frac{3\kappa^2k_3q^2 A_0^2}{2\omega^5} +\frac{(\kappa^2-4k_3^2)\kappa^2q^3 A_0^3}{2\omega^7}\Biggr]\ ,\label{I}
\ee
vanishes. Taking into account the property $(1+2xy+y^2)^{-1/2}=\sum_{n=0}^\infty P_n(-x)y^n$, where $P_n(x)$ are the Legendre polynomials, we can expand the first term of the integral around $A_0=0$ as follows 
%\be
%\frac{k_3+qA_0}{\omega_{out}}=\frac{k_3}{\omega}+\sum_{n=1}^\infty\Biggl[P_{n-1}(-\frac{k_3}{\omega})+\frac{k_3}{\omega}P_{n}(-\frac{k_3}{\omega})\Biggr]\frac{1}{\omega^n}(qA_0)^n
%\ee
\be
\frac{k_3+qA_0}{\omega_{out}}=\sum_{n=0}^\infty c_n(\vec k)(qA_0)^n\ , \nonumber
\ee
where
%\bea
%&&c_0(\vec k)=\frac{k_3}{\omega}\ , \nonumber \\
%&&c_n(\vec k)=\frac{1}{\omega^n}\Biggl[P_{n-1}\left(-\frac{k_3}{\omega}\right)+\frac{k_3}{\omega}P_{n}\left(-\frac{k_3}{\omega}\right)\Biggr] \,\, \text{for} \,\, n>0\ .
%\eea
\bea
c_0(\vec k)=\frac{k_3}{\omega}\ , \hspace{1cm}
c_n(\vec k)=\frac{1}{\omega^n}\Biggl[P_{n-1}\left(-\frac{k_3}{\omega}\right)+\frac{k_3}{\omega}P_{n}\left(-\frac{k_3}{\omega}\right)\Biggr]\,\,\,\, \,\,\,\,\,\, \text{for} \,\, n>0\ .
\eea
One can see that the first four terms of this expansion give exactly the rest of the terms of the integral \eqref{I} (the subtraction terms) with a global change of sign. Therefore they are cancelled and the integral can be written as
\be \label{I2}
I=\int_0 ^\infty k_{\perp}dk_{\perp} \sum_{n=4}^\infty\biggl[(qA_0)^n \int_{-\infty}^\infty dk_3 \,c_n(\vec k)\biggr] \, .
\ee
Under the change of variable $x=-k_3/\omega$, the integral in $k_3$ can be rewritten as
\be \label{intx}
\int_{-\infty}^\infty dk_3\,c_n(\vec k)=\frac{1}{\kappa^{n-1}}\left(\int_{-1}^1dx\,(1-x^2)^{\frac{n-3}{2}}P_{n-1}(x)-\int_{-1}^1dx\,x\,(1-x^2)^{\frac{n-3}{2}}P_{n}(x)\right) \, .
\ee
The Legendre polynomials satisfy the property $P_n(-x)=(-1)^nP_n(x)$, so it is trivial to see that for any even $n$ these integrals vanish. For odd values of $n$ and $n\ge3$ the function $(1-x^2)^{\frac{n-3}{2}}$ is a polynomial of order $n-3$. Using the property $\int_{-1}^1dx\,\text{Pol}_a(x)P_b(x)=0$ for $a<b$, where $Pol_a(x)$ is a polynomial of order $a$, we get that the integrals in \eqref{intx} vanish for $n\ge3$. This last property can be easily proven taking into account that $P_n(x)$ form a basis, and any function can be expanded as $f(x)=\sum_{b=0}^\infty c_bP_b(x)$ where $c_b=(b+1/2)\int_{-1}^{1}dx\,f(x)P_b(x)$, and if the function is a polynomial $f(x)=\text{Pol}_a(x)$, for consistency $c_b=0$ for any $b>a$. Therefore, for all values of $n$ involved in \eqref{I2} the integral vanishes, and then $I=0$, as we wanted to prove.


\begin{thebibliography}{99}
\bibitem{Heisenberg1936}
  W.~Heisenberg and H.~Euler,
  %``Consequences of Dirac's theory of positrons,''
 {\it  Z.\ Phys.} {\bf 98}, 714 (1936) 
 
 \bibitem{Sauter} F. Sauter, {\it Z. Phys.} {\bf 69}, 742 (1931). 
 
 
 
\bibitem{Schwinger51} J. Schwinger, {\it Phys. Rev. } {\bf 82}, 664 (1951). 
 

\bibitem{parker66} L. Parker, {\it The creation of particles in an expanding universe}, Ph.D. thesis, Harvard University, 1966%.  Dissexpress.umi.com, PublicationNumber  7331244
; {\it Phys.~Rev.~Lett.} {\bf 21}, 562 (1968); {\it Phys.~Rev.~D} {\bf 183}, 1057 (1969);% {\it Phys.~Rev.~D} 
 {\bf 3}, 346 (1971).


 

%\cite{Zeldovich:1970si}
\bibitem{Zeldovich1970}
  Y.~B.~Zeldovich,
  %``Particle production in cosmology,''
  {\it Pisma Zh.\ Eksp.\ Teor.\ Fiz.}  {\bf 12}, 443 (1970). 
  Y.~B.~Zeldovich and A.~A.~Starobinsky,
  %``Particle production and vacuum polarization in an anisotropic gravitational field,''
 {\it  Sov.\ Phys.\ JETP}  {\bf 34},  1159 (1972)
   [Zh.\ Eksp.\ Teor.\ Fiz.\  {\bf 61}, 2161 (1971)].
 
\bibitem{Ford1977}
  L.~H.~Ford and L.~Parker,
  %``Quantized Gravitational Wave Perturbations in Robertson-Walker Universes,''
  {\it Phys.\ Rev.\ D} {\bf 16}, 1601 (1977) 

\bibitem{hawking74} S. W. Hawking, {\it Nature} (London) {\bf 248}, 30 (1974); {\it Commun. Math. Phys.} {\bf 43}, 199 (1975). 
   
  
\bibitem{parker-toms}L.~Parker and D.~J.~Toms, {\it Quantum Field Theory in Curved Spacetime: Quantized Fields
and Gravity} (Cambridge University Press, Cambridge, England 2009).
\bibitem{Waldbook} R.~M.~Wald, {\it Quantum Field Theory in Curved Space-time and Black Hole Thermodynamics} (University of Chicago Press, Chicago, 1994).
\bibitem{fulling} S.~Fulling, {\it Aspects of Quantum Field Theory in Curved Space-Time} (Cambridge University Press, Cambridge, England, 1989).



\bibitem{birrell-davies} N.~D.~Birrell  and P.~C.~W.~Davies, {\it Quantum Fields in Curved Space} (Cambridge University Press, Cambridge, England, 1982).

\bibitem{narozhnyi1970}
  N.~B.~Narozhnyi and A.~I.~Nikishov,
  %``The Simplist processes in the pair creating electric field,''
 {\it Yad.\ Fiz.}  {\bf 11}, 1072 (1970)
   [Sov.\ J.\ Nucl.\ Phys.\  {\bf 11} 596 (1970)].
   
\bibitem{Brezin-Itzykson} E. Brezin and C. Itzykson, {\it Phys. Rev. D} {\bf 2},  1191 (1970).   
   


\bibitem{parker-fulling} L.~Parker and S.~A.~Fulling, {\it Phys.~Rev.~D} {\bf 9}, 341 (1974); S. A. Fulling and L. Parker, {\it Ann.~Phys.} (N.Y.) {\bf 87}, 176 (1974).
\bibitem{Birrell78} N. D. Birrell,  {\it Proc. R. Soc.  B} {\bf 361}, 513 (1978).

\bibitem{Anderson-Parker}  P.~R.~Anderson and L.~Parker, {\it Phys.~Rev.~D} {\bf 36}, 2963 (1987).


\bibitem{ANOP}  I.~Agullo, J.~Navarro-Salas, G.~J.~Olmo, and  L.~Parker, {\it Phys.~Rev.~Lett.} {\bf 103}, 061301 (2009);
{\it Phys.~Rev.~D} {\bf 81}, 043514, (2010); I. Agullo, W. Nelson, and A. Ashtekar {\it Phys. Rev. D} {\bf 91},  064051 (2015). 

\bibitem{delRio-Navarro15} A. del Rio and J. Navarro-Salas, {\it Phys. Rev. D} {\bf 91}, 064031 (2015).

%A. Landete, J. Navarro-Salas and F. Torrenti, {\it Phys. Rev. D} {\bf 89} 044030 (2014); S. Ghosh, {\it Phys. Rev. D} {\bf 91}, 124075 (2015); ; S. Ghosh, {\it Eur. Phys. J. C} {\bf 79}, 239 (2019);
%\bibitem{parker2012} L. Parker, {\it J.\ Phys.\ A}  {\bf 45},  374023 (2012).


\bibitem{rio1} A. Landete, J. Navarro-Salas, and F. Torrenti, {\it Phys. Rev. D} {\bf 88}, 061501 (2013); {\bf 89} 044030 (2014); A. del Rio, J. Navarro-Salas, and F. Torrenti, {\it Phys. Rev. D} {\bf 90}, 084017 (2014); S.~Ghosh, {\it Phys.~Rev.~D} {\bf 91}, 124075 (2015); {\bf 93},   044032 (2016); A. del Rio, A. Ferreiro, J. Navarro-Salas, and F. Torrenti {\it Phys. Rev. D} {\bf 95}, 105003 (2017). 

\bibitem{BFNV}J. F. Barbero G., A. Ferreiro, J. Navarro-Salas, and E. J. S. Villase\~nor, {\it Phys. Rev. D} {\bf 98}, 025016 (2018).




\bibitem{ELI}The extreme light infrastructure (ELI) project, www.extreme-light-infrastructure.eu/ .

\bibitem{Yakimenko19} V. Yakimenko et al.,  {\it Phys. Rev. Lett.} {\bf 122}, 190404 (2019). 

%schwinger effect in cosmological scenarios

\bibitem{R} R. Ruffini, G. Vereshchagin, and S. Xue, {\it Phys. Rep.} {\bf 487}, 1 (2010).

\bibitem{Kim:2019joy}
  S.~P.~Kim,
 % {\it Astrophysics in Strong Electromagnetic Fields and Laboratory Astrophysics}, 
  arXiv:1905.13439.% [gr-qc]

\bibitem{Cosmo1}M. B. Fr\"ob, J. Garriga, S. Kann, M. Sasaki, J. Soda, T. Tanaka, and A. Vilenkin, {\it J. Cosmol. Astropart. Phys.} 04 (2014) 009; T. Kobayashi and N. Afshordi, {\it J. High Energy Phys.} {\bf 10}, (2014) 166; C. Stahl, E. Strobel, and S.-S. Xue, {\it Phys. Rev. D} {\bf 93}, 025004 (2016).


\bibitem{Scahl}C.~Stahl,
  %``Schwinger effect impacting primordial magnetogenesis,''
  {\it Nucl.\ Phys.} {\bf B939}, 95 (2019).

\bibitem{SGF19} S.~Shakeri, M.~A.~Gorji, and H.~Firouzjahi,
  %``Schwinger Mechanism During Inflation,''
 {\it  Phys.\ Rev.\ D} {\bf 99}, 103525 (2019).



%{\color{blue}\bibitem{PSpliting} T. Hayashinaka and J. Yokoyama, JCAP07(2016)012.}




%\bibitem{reheating}  L. A.  Kofman, A. D. Lindle, and A. A. Starobinsky, {\it Phys. Rev. Lett.} {\bf 73}, 3195 (1994); {\it Phys. Rev. Lett.}  {\bf 76}, 1011 (1996).
 %J. Baacka, K. Heitmann and C. Ptzold,  {\it Phys~Rev.~D} {\bf 58}, 125013 (1998); P. Greene and L. A. Kofman,  {\it Phys. Lett.}{\bf B}448, 6 (1999) .


%\bibitem{Mueller} N. Mueller, F. Hebenstreit, and J. Berges {\it Phys. Rev. Lett.} {\bf 117}, 061601 (2016); S. P. Kim and H. K. Lee {\it Phys. Rev. D} {\bf 76}, 125002 (2007).

\bibitem{Dunne} G. V. Dunne, {\it Int. J. Mod. Phys. A} {\bf 27}, 1260004 (2012); {\it  Eur. Phys. J. D} {\bf 55}, 327 (2009); B. S. Xie, Z. L. Lie, and S. Tang, {\it Matter Radiat. Extremes} {\bf 2}, 225 (2017).


%\bibitem{qftbook} M.E. Peskin and D. V. Schroeder, {\it An Introduction to Quantum Field Theory}, Addison-Wesley, Reading, MA, USA (1995).

%\bibitem{Israel} W. Israel, {\it Phys. Rev. Lett.} {\bf 57}, 397 (1986).

%\bibitem{Hawking} S. W. Hawking, {\it Black Holes aren't Black}, Gravity Research Foundation (1974), essay [www.gravityresearchfoundation
%738 .org/year\#1974]; {\it Commun. Math. Phys.} {\bf 43}, 199 (1975).

 
   \bibitem{Dunne2}R. Dabrowski and G. V. Dunne {\it Phys. Rev.D} {\bf 94} 065005 (2016).
  

  
\bibitem{Garay} L. J. Garay, A. Garc\'ia Mart\'in-Carom and M. Mart\'in-Benito, % \textit{Unitary quantization of a charged scalar field and Schwinger effect},  arXiv:1911.03205v1
{\it J. High Energy Phys.} 04 (2020) 120.

\bibitem{BFNP} P. Beltr\'an, A. Ferreiro, J. Navarro-Salas, and S. Pla,
 {\it Phys. Rev. D} {\bf 100}, 085014 (2019).




 \bibitem{Cooper-Mottola89} F. Cooper and E. Mottola, {\it Phys. Rev. D} {\bf 40}, 456 (1989). 
\bibitem{Kluger91} Y. Kluger, J. M. Eisenberg, B. Svetitsky, F. Cooper and E. Mottola, {\it Phys. Rev. Lett.} {\bf 67}, 2427 (1991).
\bibitem{Kluger92}Y. Kluger, J. M. Eisenberg, B. Svetitsky, F. Cooper and E. Mottola, {\it Phys. Rev. D} {\bf 45}, 4659 (1992).

\bibitem{FN} A. Ferreiro and J. Navarro-Salas, {\it Phys. Rev. D} {\bf 97}, 125012   (2018).

\bibitem{FNP} A. Ferreiro, J. Navarro-Salas, and S. Pla, {\it Phys. Rev. D} {\bf 98}, 045015 (2018).

\bibitem{FNP2}  A. Ferreiro, J. Navarro-Salas, and S. Pla, % \textit{Pair creation in electric fields, renormalization, and backreaction}, 
arXiv:1903.11425. %To appear in the Proceedings of the 15$^{th}$ Marcel Grossmann Meeting, Rome (2018). 







 
%\bibitem{mathbook} M. Abramowitz, and I. A. Stegun, {\it Handbook of Mathematical Functions: With Formulas, Graphs and Mathematical Tables}, Dover, New York (1972).




%Y. Kluger, J. M. Eisenberg, B. Svetitsky, F. Cooper and E. Mottola, {\it Phys. Rev. D} {\bf 45}, 4659 (1992).



%\bibitem{BFNP} P. Beltr\'an, A. Ferreiro, J. Navarro-Salas and S. Pla {\it Phys. Rev. D} {\bf 100}, 085014 (2019)

%\bibitem{BFNV}  %Non-equilibrium proceses
\bibitem{BNP} P. Beltr\'an-Palau, J. Navarro-Salas, and S. Pla,  {\it Phys. Rev. D} {\bf 99} 105008 (2019).


\bibitem{Had1}V. Balakumar, E. Winstanley, {\it Classical and Quantum Gravity} {\bf 37}, 065004 (2020).% \textit{Hadamard renormalization for a charged scalar field},  arXiv:1910.03666v1.

\bibitem{PointS} R. Herman and W. A. Hiscock,  {\it Phys. Rev. D} {\bf 53}, 3285 (1996).

 \bibitem{DeWittbook} B. S. DeWitt, {\it Dynamical theory of groups and fields} (Gordon and Breach, New York 1965). 


\bibitem{Gilkey} P. B. Gilkey, {\it J. Differ. Geom. } {\bf 10}, 601 (1975).

\bibitem{Vassilevich} D. V. Vassilevich, {\it Phys. Rep.} {\bf  388}, 279 (2003).

%{\color{blue} \bibitem{FNP20} A. Ferreiro, J. Navarro-Salas and S. Pla, \textit{$R$-summed form of adiabatic expansions in curved spacetime,}    arXiv:2003.09610. }
%\bibitem{Mueller} N. Mueller, F. Hebenstreit, and J.Berges {\it Phys. Rev. Lett.} {\bf 117}, 061601 (2016).
%\bibitem{thooft1} G.'t Hooft, {\it Phys. Rev. Lett.} {\bf 37} 8 (1976) .
%\bibitem{Christ} N. H. Christ, {\it Phys. Rev. D} {\bf21}1591 (1980) .

%Quantum Field Theory in Curved Spacetime



\bibitem{Pittrich-Gies} W. Pittrich and H. Gies, {\it Probing the Quantum Vacuum}, Springer, Heidelberg (2000).


\bibitem{duff} M. J. Duff, {\it Classical  Quantum Gravity} {\bf 11}, 1387 (1994).

%\bibitem{anomaly1} J.F. Donoghue and K. El-Menoufi, {\it J. High Energ. Phys}. {\bf 05}, 118 (2015) 

\bibitem{mu}A. Ferreiro and J. Navarro-Salas, {\it Phys. Lett. B} {\bf 792}, 81 (2019). 


\bibitem{qftbook} M.E. Peskin and D. V. Schroeder, {\it An Introduction to Quantum Field Theory} (Addison-Wesley, Reading, MA, 1995).



%S. A. Fulling, L. Parker and B. L. Hu, {\it Phys. Rev. D} {\bf 10}, 3905 (1974). T.~S.~Bunch, {\it J.~Phys.~A} {\bf13}, 1297 (1980). P.~R.~Anderson and L.~Parker, {\it Phys.~Rev.~D} {\bf 36}, 2963 (1987)


%\bibitem{Bunch80} T.~S.~Bunch, {\it J.~Phys.~A} {\bf13}, 1297 (1980). 




%\bibitem{parker68} L.~Parker, {\it Phys.~Rev.~Lett.} {\bf 21}, 562 (1968); {\it Phys.~Rev.~D} {\bf 183}, 1057 (1969); {\it Phys.~Rev.~D} {\bf 3}, 346 (1971).


%\bibitem {Anderson-Parker} 
%P.~R.~Anderson and L.~Parker, {\it Phys.~Rev.~D} {\bf 36}, 2963 (1987).  
%P.~R.~Anderson and W.~Eaker, {\it Phys. Rev. D} {\bf 61}, 024003 (1999). S.~Habib, C.~Molina-Paris   and E.~Mottola, {\it Phys.~Rev.~D} {\bf 61}, 024010 (1999).  I.~Agullo, J.~Navarro-Salas, G.~J.~Olmo and  L.~Parker, {\it Phys.~Rev.~Lett.} {\bf 103}, 061301 (2009);{\it Phys.~Rev.~D} {\bf 81}, 043514, (2010). I. Agullo, A. Landete and J. Navarro-Salas, {\it Phys. Rev. D} {\bf90}, 124067 (2014). A.~del Rio and  J.~Navarro-Salas, {\it Phys.~Rev.~D} {\bf 91}, 064031 (2015)


%I.~Agullo, J.~Navarro-Salas, G.~J.~Olmo and  L.~Parker, {\it Phys.~Rev.~Lett.} {\bf 103}, 061301 (2009);
%{\it Phys.~Rev.~D} {\bf 81}, 043514, (2010).
%\bibitem{Ghosh16} S.~Ghosh, {\it Phys.~Rev.~D} {\bf 91}, 124075 (2015); {\it Phys.~Rev.~D} {\bf 93},   044032 (2016).
%\bibitem{RFNT} A.~del Rio, A. Ferreiro, J.~Navarro-Salas and F.~Torrenti, {\it Phys.~Rev.~D} {\bf 95}, 105003 (2017).


%\bibitem{CGHS} C. G. Callan, S. B. Giddings, J. A. Harvey and A. Strominger, {\it Phys. Rev. D} {\bf 45}, 1005 (1992). J. A. Harvey and A. Strominger, {\it Quantum aspects of black holes}, arXiv:hep-th/9209055.
%\bibitem{FabbriNavarro} A. Fabbri and J. Navarro-Salas, {\it Modeling black hole evaporation}, ICP-World Scientific, London (2005).



%\bibitem{Schwinger62} J. Schwinger, {\it Phys. Rev.} {\bf 128}, 2425 (1962).

%\bibitem{Bertlmann} R. A. Bertlmann, {\it Anomalies in Quantum Field Theory}, Oxford University Press, Oxford (2000).

%\bibitem{peskin-schroeder} M. E. Peskin and D. V. Schroeder, {\it Introduction to Quantum Field Theory},  Addison-Wesley, Reading, 1995
%\bibitem{Searlyuniverse} M. B. Froöb, J. Garriga, S. Kann, M. Sasaki, J. Soda, T. Tanaka, A. Vilenkin, {\it J. Cosmol. Astropart. Phys.} {\bf 1404}, 009 (2014).  T. Kobayashi and N. Afshordi, {\it J. High Energy Phys.} {\bf 10}, 166 (2014).  C. Stahl and S.-S. Xue, {\it Phys. Rev. b} {\bf 760}, 288 (2016). C. Stahl, E. Strobel and S.-S. Xue, {\it Phys. Rev. D} {\bf 93}, 025004 (2016). C. Stahl, {\it Schwinger effect impacting primordial magnetogenesis}, arXiv:1806.06692 

%\bibitem{Ivan} I. Agullo, W. Nelson and A. Ashtekar {\it Phys. Rev. D} {\bf 91},  064051 (2015).





%\bibitem{Cooper1}F. Cooper and E. Mottola, {\it Phys. Rev. D} {\bf 40}, 456 (1989).
%\bibitem{Kluger92} Y. Kluger, J. M. Eisenberg, B. Svetitsky, F. Cooper and E. Mottola, {\it Phys. Rev. D} {\bf 45}, 4659 (1992).
%\bibitem{Kluger93} Y. Kluger, J. M. Eisenberg and B. Svetitsky,  {\it Int. J. Mod. Phys. E} {\bf 2}, 333 (1993).

%\bibitem{CF} S. M. Christensen and S. A. Fulling, {\it Phys. Rev. D}{\bf 15}, 2088 (1977).
%\bibitem{ADNS} I. Agullo, A. del Rio and J. Navarro-Salas, {\it Int. J. Mod. Phys. D} {\bf 26}, 1742001 (2017); {\it Phys. Rev. Lett.} {\bf118}, 111301 (2017).


%\bibitem{BFNV} J. F. Barbero, A. Ferreiro, J. Navarro-Salas and E.J.S. Villase\~nor, {\it Adiabatic expansion for Dirac fields, renormalization, and anomalies}, arXiv:1805.05107.



%%schwinger effect in cosmological scenarios
%{\color{red}\bibitem{Frob}M. B. Fr\"ob, J. Garriga, S. Kann, M. Sasaki, J. Soda, T. Tanaka, A. Vilenkin, {\it J. Cosmol. Astropart. Phys.} 1404 (2014) 009
%\bibitem{KA} T. Kobayashi and N. Afshordi, {\it J. High Energy Phys.} {\bf 10}, 166 (2014).
%%\bibitem{SX} C. Stahl and S.-S. Xue, {\it Phys. Rev. b} {\bf 760}, (2016) 288-292
%\bibitem{SSX} C. Stahl, E. Strobel and S.-S. Xue, {\it Phys. Rev. D} {\bf 93}, 025004 (2016).
%\bibitem{PSpliting} T. Hayashinaka and J. Yokoyama, JCAP07(2016)012.}

%\bibitem{Waldbook} R.~M.~Wald, {\it Quantum Field Theory in Curved Space-time and Black Hole Thermodynamics}, University of Chicago Press, Chicago, (1994).
%\bibitem{fulling} S.~Fulling, {\it Aspects of Quantum Field Theory in Curved Space-Time}, Cambridge University Press, Cambridge, England (1989).

%\bibitem{fulling73} S. A. Fulling, {\it Phys. Rev. D} {\bf 7}, 2850 (1973).



%\bibitem{ford-parker} L. H. Ford and L. Parker, {\it Phys. Rev. D} {\bf 16}, 1601 (1977).

%\bibitem{Grishchuk} L.P. Grishchuk, {\it Sov. Phys. JETP} {\bf 40}, 409 (1975).

%\bibitem{guth} A. H. Guth {\it Phys.Rev. D} {\bf 23}, 347 (1981).
%\bibitem{inflation2} Mukhanov V.F. and Chibisov G.V., {\it JETP Letters} {\bf33}, 532(1981).
%
%Hawking S. W., {\it Phys. Lett.} B{\bf 115}, 295 (1982).
%
%Guth A. and Pi S.-Y., {\it Phys. Rev. Lett.} {\bf 49}, 1110 (1982).
%
%Starobinsky A. A., {\it Phys. Lett.} B{\bf 117}, 175 (1982).
%
%Bardeen J.M., Steinhardt P.J. and Turner M.S., {\it Phys. Rev.} D {\bf 28}, 679 (1983).

%\bibitem{KLS} L.~Kofman, A.~Linde and A.~Starobinsky, {\it Phys.~Rev.~Lett.}  {\bf 73}, 3195 (1994); {\it Phys.~Rev.~D} {\bf 56}, 3258 (1997).

%Schwinger Effect


%\bibitem{Heisenberg-Euler} W. Heisenberg and H. Euler, {\it Z. Phys.} {\bf 98}, 714 (1936).
%\bibitem{Dunne0} G. V. Dunne, G. V. Dunne, {\it Int.\ J.\ Mod.\ Phys.\ A} {\bf 27}, 1260004 (2012); {\it  Eur. Phys. J. D} {\bf 55}, 327 (2009). B.S. Xie, Z. L. Lie and S. Tang, {\it Matter and Radiation at Extremes},  {\bf 2}, 225 (2017).

%Adiabatic Regularization for gravitational field


%Adiabatic Regularization for electric field
%\bibitem{FN} A. Ferreiro and J. Navarro-Salas, {\it Pair creation in electric fields, anomalies, and renormalization of the electric current}, e-Print: arXiv:1803.03247
%\bibitem{duff} M. J. Duff, {\it Nuclear Physics  B} {\bf 125}, 334-348 (1977)




%initial conditions 





%regularization electric field in de sitter
%\bibitem{anderson-mottola}P. Anderson and E. Mottola {\it Phys. Rev.  D} {\bf 89}, 104039 (2014)
%anomalies
%\bibitem{arxiv} J. F. Donoghue and B. elK. El-Menoufi, {\it JHEP} {\bf 1505}, 118 (2015).
%\bibitem{duff2} M. J. Duff, {\it Nuclear Physics B} {\bf 125}, 334-348 (1977)
%\bibitem{davies} P. C. W. Davies and W. G. Unruh {\it Proc. R. Soc. Lond. A} {\bf 356}, 259-268 (1977)
%\bibitem{Duff} M.~J.~Duff, {\it Class. Quantum Grav.} {\bf 11}, 1387 (1994).

%otros


%\bibitem{Blaer} A. S. Blaer, N. H. Christ, and J-F. Tang, {\it Phys. Rev. Lett.} {\bf 47}, 1364 (1981).







%{\color{red}\bibitem{FabbriNavarro} A. Fabbri and J. Navarro-Salas, {\it Modeling Black Hole Evaporation}, ICP-World Scientific, London (2005). NO LA CITAMOS!!}

%\bibitem{Brezin-Itzykson} {\color{blue}E. Brezin and C. Itzykson, {\it Phys. Rev. D} {\bf 2}, 1191 (1970).}







\end{thebibliography}
\end{document}